\documentclass[amsmath,amssymb,aps, pra, reprint,
citeautoscript, superscriptaddress]{revtex4-2}

\usepackage{dcolumn}
\usepackage[utf8]{inputenc}
\usepackage[T1]{fontenc}
\usepackage{mathptmx}
\usepackage{etoolbox}
\usepackage{xcolor}
\usepackage{epsfig,graphicx,amssymb,color,amssymb,amsmath,mathrsfs,bm,bbold}
\usepackage[abs]{overpic} 

\makeatother
\newcommand{\ha}{\hat{a}}
\newcommand{\hb}{\hat{b}}
\newcommand{\hsp}{\hat{\sigma}^{+}}
\newcommand{\hsm}{\hat{\sigma}^{-}}
\newcommand{\hsx}{\hat{\sigma}^{x}}
\newcommand{\hsy}{\hat{\sigma}^{y}}
\newcommand{\hsz}{\hat{\sigma}^{z}}

\newcommand{\bsz}{\bar{\sigma}_{z}}

\pdfoutput=1

\begin{document}
\include{commands}

\title{Scattering theory of frequency-entangled biphoton states facilitated by cavity polaritons}

\author{Andrei Piryatinski}
\email{apiryat@lanl.gov}
\affiliation{Theoretical Division, Los Alamos National Laboratory, Los Alamos, NM 87545, United~States }

\author{Nishaant Jacobus}
\affiliation{Theoretical Division, Los Alamos National Laboratory, Los Alamos, NM 87545, United~States }
\affiliation{Chemical Physics Theory Group, Department of Chemistry, and Center for Quantum Information and Quantum Control,
University of Toronto, Toronto, Ontario M5S 3H6, Canada}

\author{Sameer Dambal}
\affiliation{Theoretical Division, Los Alamos National Laboratory, Los Alamos, NM 87545, United~States }
\affiliation{Department of Physics, University of Houston, Houston, Texas 77204, United~States}

\author{Eric~R.~Bittner}
\email{ebittner@central.uh.edu}
\affiliation{Department of Physics, University of Houston, Houston, Texas 77204, United~States}

\author{Yu Zhang}
\affiliation{Theoretical Division, Los Alamos National Laboratory, Los Alamos, NM 87545, United~States }

\author{Ajay Ram Srimath Kandada}
\affiliation{Department of Physics and Center for Functional Materials, Wake Forest University, 1834 Wake Forest Road, Winston-Salem,
NC 27109, United States}

\begin{abstract}
The use of quantum light to probe exciton properties in semiconductor and molecular nanostructures typically occurs in the low-intensity regime. A substantial enhancement of exciton-photon coupling can be achieved with photonic cavities, where excitons hybridize with cavity modes to form polariton states. To provide a theoretical framework for interpreting emerging experimental efforts in this direction, we develop a scattering theory describing the interaction of frequency-entangled photon pairs with cavity polariton and bipolariton states under various coupling regimes. Employing the Tavis-Cummings model in combination with our scattering approach, we present a quantitative analysis of how the interaction of the entangled photon pair with the polariton/bipolariton modifies its joint spectral amplitude (JSA). Specifically, we examine the effects of the cavity-mode steady-state population, exciton-cavity coupling strength, and different forms of the input photon JSA. Our results show that the entanglement entropy of the scattered photons is highly sensitive to the interplay between the input JSA and the spectral line shapes of the polariton resonances, emphasizing the cavity filtering effects. We suggest that biphoton scattering quantum light spectroscopy best serves as a sensitive probe of polariton and bipolariton states in the photon-vacuum cavity  state. {\color{black} Our approach is not only robust to various regimes of cavity-exciton coupling, but also amenable to extensions beyond the Tavis-Cummings model, enabling the representation of a broad class of molecular systems and solid state quantum materials.}

\end{abstract}

\date{\today}
 
\maketitle



\begin{figure*}[t!]
\centering
\includegraphics[width=0.85\textwidth]{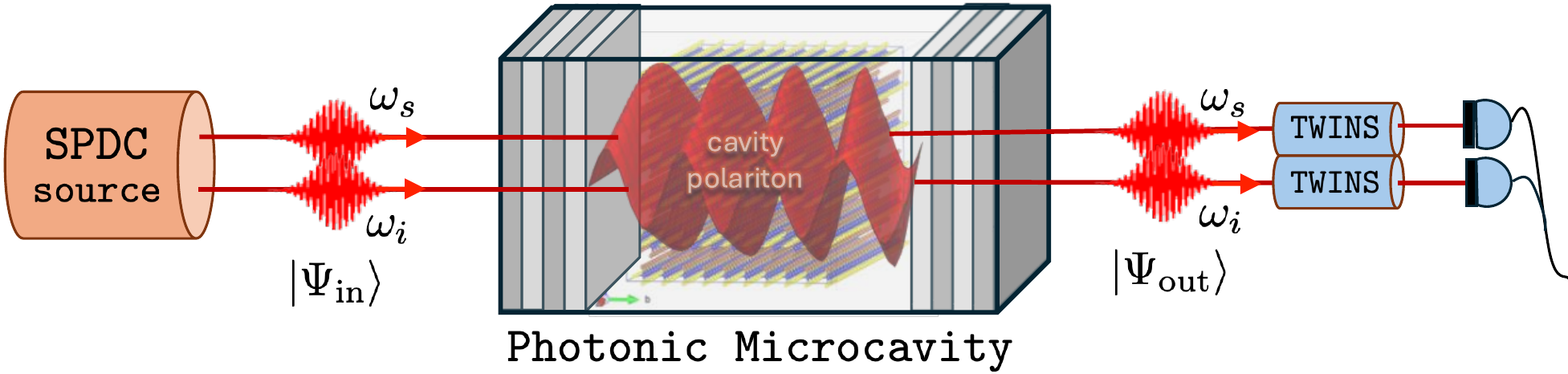}
\caption{Schematics of a photonic microcavity containing QEs, e.g., semiconductor nanostructure, where the interaction between a cavity photon mode and nanostructured material excitons forms a hybrid light–matter quasiparticle — the cavity polariton. {\color{black}The polariton steady state can be prepared by applying a weak incoherent pump to the QEs (not shown in the plot), which also results in a steady-state cavity photon population.} The polariton state is further probed by scattering an input entangled photon pair, $|\Psi_\text{in}\rangle$, generated via spontaneous parametric down-conversion (SPDC), with the output state, $|\Psi_\text{out}\rangle$, detected after transmission through the TWINS interferometric setup.}
\label{Fig:Schm}
\end{figure*}

\section{Introduction}

Quantum light has recently been shown, both theoretically and experimentally, to be a powerful probe of electronic and vibrational dynamics in materials~\cite{Dorfman_RMP:2016,MukamelJPhysB:2020,Szoke_JMaterChemC:2020,Eshun_AccChemRes:2022}.  This sensitivity arises from the unique characteristics of quantum photon states, such as entanglement and squeezing, and from the ability to detect changes in nonclassical photon statistics induced by the material response. {\color{black} Early experiments showed that entangled photons can  enhance the two-photon absorption cross section, enabling low-intensity probing of molecular systems~\cite{Lee_JPCB:2006}. Subsequent theoretical work developed frameworks for describing nonlinear interactions between entangled photons and molecular systems and elucidated the physical mechanisms underlying this enhancement, including nonclassical intensity fluctuations~\cite{Schlawin_JPhysB:2017,Schlawin_AccCheRes:2018,RaymerJCP:2021}. More recent extensions include theoretical study  of  the two-photon absorption due to the interaction with broadband squeezed vacuum~\cite{Raymer_PRA:2022} and high-intensity pulsed entangled beams~\cite{Schlawin_JCP:2024}.
 
Capitalizing on the enhanced sensitivity provided by two-photon absorption probes, entangled photon pulse sequences have been considered to interrogate excited state dynamics in molecular systems giving rise to quantum light two-dimensional (2D) spectroscopies. Theoretical studies include quantum-photon four-wave-mixing processes~\cite{Richter_PRA:2010} and entangled-photon 2D fluorescence spectroscopy~\cite{Raymer_JPCB:2013}. Related approaches involve entangled pump–probe techniques based on two-photon coincidence detection~\cite{Fujihashi_JCP:2024,Fujihashi_arXiv:2025}, as well as entangled three-photon techniques~\cite{Fujihashi_JCP:2021}. Theoretical foundation for coherent quantum 2D spectroscopies has also been proposed~\cite{Ishizaki_JCP:2020}. Furthermore, quantum interferometric schemes employing two parametric down-conversion sources have been proposed to selectively probe exciton–exciton interactions in molecular aggregates~\cite{Kizmann_PNAS:2023}. Finally, pathway selectivity enabled by 2D techniques has been proposed for electronic–vibrational spectroscopy using entangled visible and infrared photons~\cite{Jadoun_LasPhotRev:2025}, and the interplay between entangled and unentangled photon contributions to the 2D signal in sequences of squeezed vacuum beams has been elucidated~\cite{Jadoun_PRL:2025}.

}

Experimental studies have shown that quantum interference of entangled photons in a Hong–Ou–Mandel (HOM) interferometer, with a sample introduced into one of the interferometer arms, enables the extraction of ultrafast dephasing times on the femtosecond scale using a continuous-wave laser source~\cite{Kalashnikov:SciRep:2017}. This work has stimulated further studies showing that HOM interferometry with entangled photons holds potential for probing ultrafast dynamics of elementary excitations in condensed matter systems~\cite{Li_QSciTec:2018} and can  be used to enhance the accuracy of light–matter interaction measurements in molecular condensed matter systems~\cite{Dorfman_CommPhys:2021,Eshun_JACS:2021,Arango_JPCA:2023,Arango_JPCA:2024}.

Another interferometric configuration offering a spectroscopically viable strategy to measure the joint spectral amplitude (JSA) of a frequency entangled photon pair has been proposed in Ref.~\cite{Moretti_JCP:2023}. In this approach, two Fourier Transform spectrometers are inserted into the signal and idler channels prior to photon detection. This method employs a Translating-Wedge Interferometry (TWINS) approach, where a single photon is split into two replicas via polarization control, and a variable delay is introduced using birefringent wedges. Detection as a function of delay yields an interferogram, whose Fourier transform reveals the spectral content. For entangled photon pairs, independent TWINS modules are applied to each photon, and the joint interferogram—constructed from coincidence counts—provides access to the joint spectral amplitude (JSA) via two-dimensional Fourier analysis. Further analysis of JSA enables direct extraction of the entanglement entropy of the photon pair before and after interaction with the sample. Furthermore, the use of a single TWINS interferometer placed in the signal channel has been shown to provide simultaneous time- and frequency-resolved quantum-photon spectroscopic probe~\cite{Alvarez_NatComm:2025}.

Optical microcavities enable strong light–matter interactions that are particularly useful platforms for low-intensity quantum-photon spectroscopies. Such interactions give rise to hybrid light–matter quasiparticles known as polaritons, formed via the coupling between cavity photons and elementary material excitations. Recent experimental studies have investigated the propagation of frequency-entangled photons through an empty photonic cavity, revealing deviations from the conventional optical filtering behavior~\cite{Malatesta_arXiv:2023}. Introducing strongly coupled quantum emitters (QEs) into the cavity will bring further complexity in understanding the polariton effects on the quantum photon JSA and entanglement entropy. 

Given recent advances in polariton chemistry~\cite{Schafer_NatCom:2022,Schwennicke_ChemSocRev:2025,Yin_Science:2025}, quantum-light spectroscopy offers promising opportunities {\color{black} for probing molecular cavity polaritons~\cite{Debnath_JAP:2020,Debnath_FrontPhys:2022}. } At the same time, emerging studies on the unique properties of cavity quantum materials~\cite{SchlawinAPR:2022} open new perspectives for applying quantum-light techniques as sensitive probes of strongly correlated electronic, vibrational, orbital, and spin degrees of freedom. Advancing the use of quantum-light techniques in this context calls for the development of theoretical frameworks capable of describing the scattering of entangled photon states by cavity polaritons in strongly correlated condensed-matter systems.

In most of the interferometric setups discussed above the sample of interest is placed into one of the interferometer arms or into either signal or idler channel. However, the use of photonic cavity in combination with the TWINS interferometers allows one to send both signal and idler photons through the cavity containing sample as illustrated Fig.~\ref{Fig:Schm}. In this case, the correlations within the bipolaritons can result in changes in the photon entanglement measure and subsequently lead to spectroscopic signatures of these correlations. {\color{black} Identification of these spectroscopic signatures and further development of experimental techniques require the development of theoretical approaches capable of treating material excitations interacting with cavity modes and scattered quantum photons on the same footing. }

To date this theoretical work focused on identifying the effects of fluctuations and dissipation in correlated polariton states~\cite{Li_JCP:2019}. The authors developed a frequency-domain perturbative approach for the biphoton scattering matrix to identify the role of phase fluctuations. In addition, a time-domain polariton scattering approach based on the Kubo-Anderson-type second cumulant expansion was employed to demonstrate that correlations among polariton bath fluctuations can give rise to photon entanglement entropy production. Employing a molecular dimer model, the role of  exciton–exciton interactions on the entanglement entropy of a probing biphoton pair was studied in detail, revealing their spectroscopic signatures~\cite{Bittner_JCP:2020}. A further generalization of this model, incorporating dissipation processes via the Lindblad master equation, was subsequently proposed, investigating the interplay between photonic and excitonic subsystem~\cite{Giri_JCP:2025}.

The evolution of the JSA arising from biphoton scattering mediated by cavity polaritons has been investigated in Ref.~\cite{wgnx-l7qx}. Utilizing the Tavis–Cummings model to describe the polariton states and employing the Møller scattering operator formalism, the authors implemented a Gaussian-preserving map between the input and output biphoton JSAs. This approach provided a computationally efficient framework for predicting the output JSA given an experimentally measured form of the input JSA, albeit at the expense of neglecting polariton dissipation processes responsible for line broadening. In this work, we propose a generalized scattering-matrix framework that establishes a direct connection between the input and output biphoton JSA, while consistently accounting for both the coherent and dissipative dynamics of the polariton system.

Below, we assume the spectroscopic setup illustrated in Fig.~\ref{Fig:Schm}, where an entangled photon pair produced via a spontaneous parametric downconversion (SPDC) process passes through the cavity. Both photons in the signal and idler channels participate in the scattering process facilitated by cavity single- and bipolariton states. The JSA of the scattered photons is further extracted through the aforementioned TWINS interferometry setup proposed and used in  Refs.~\cite{Moretti_JCP:2023,Malatesta_arXiv:2023}. In Sec.~\ref{Sec:S-matrix}, we derive the biphoton scattering operator in terms of single- and bipolariton Green functions and discuss the contributions of various scattering amplitudes. The relationship between the input and output JSAs mediated by these scattering amplitudes is established in Sec.~\ref{Sec:JSA-scat}. The specific implementation of these relationships for cavity polaritons described by the Tavis–Cummings model is presented in Sec.~\ref{Sec:TCJSA}, and the corresponding results are discussed in Sec.~\ref{Sec:Res}. Sec.~\ref{Sec:Conc} concludes the paper.

\section{Biphoton scattering matrix}
\label{Sec:S-matrix}

\begin{figure*}[t!]
\centering
\includegraphics[width=0.85\textwidth]{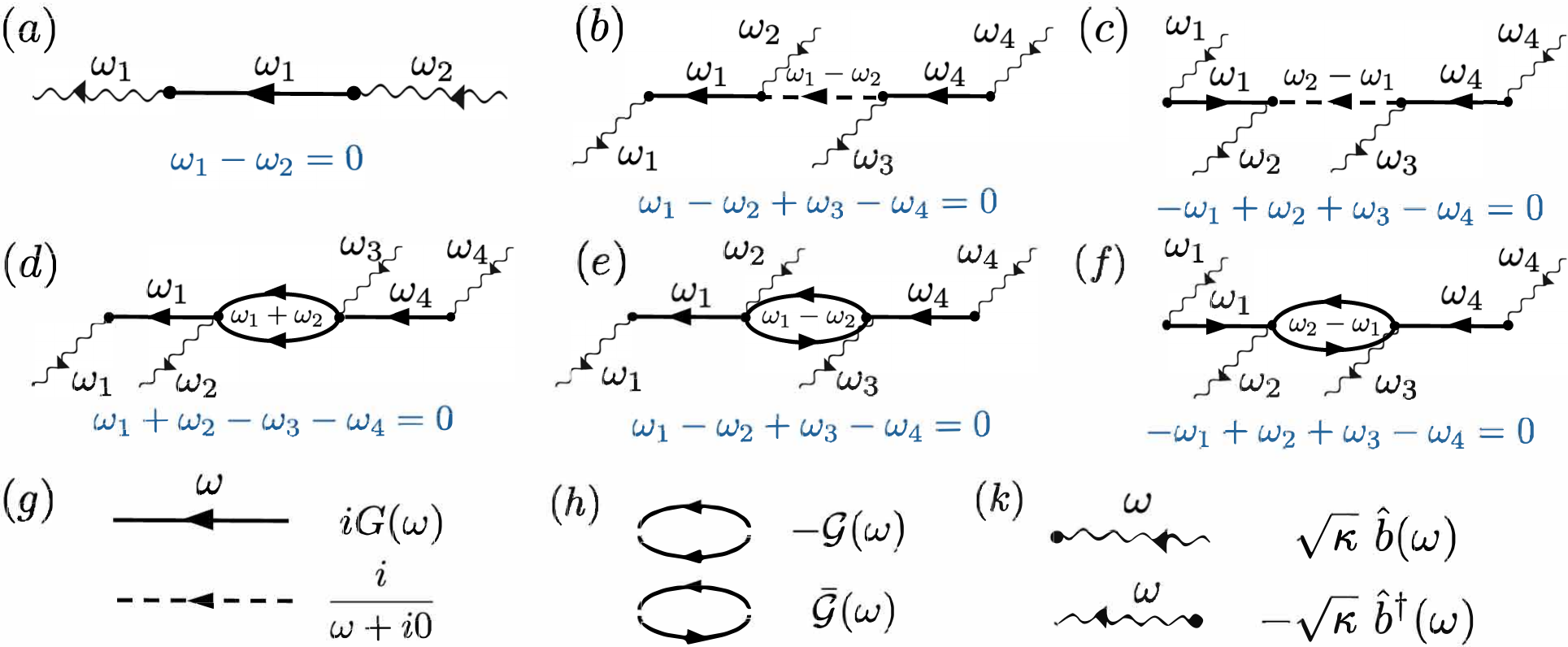}
\caption{Feynman diagrams describing (a) single-photon scattering by the polariton mode and (b, c) biphoton scattering involving the steady-state propagation. The other diagrams describe the biphoton scattering events that involve (d) the bipolariton coherence and (e,f) the polariton-polariton excited state coherences. The total energy of the incoming and scattered photons during the scattering is conserved, as indicated in the blue caption below each diagram. The diagram correspondence rules are shown in (g) for single-polariton and the steady-state Green functions, (h) for the bipolariton and  polariton-polariton coherence Green functions, and (k) for incoming and scattered photon lines with the vertex, respectively. The corresponding expression for each diagram (a)-(f) should by multiplied by the prefactor $2\pi\delta(\omega)$, where $\omega$ is substituted with the associated energy conservation expression and integrated over all frequency variables.}
\label{Fig:FD1}
\end{figure*}

Let us introduce a photonic cavity model that supports a single photon mode with frequency $\omega_c$, described by the creation operator $\ha^\dag$ and annihilation operator $\ha$ satisfying the commutation relation $[\ha,\ha^\dag]=1$. The cavity environment consists of a photon continuum. Each mode of this continuum falling in the frequency interval between $\omega$ and $\omega+d\omega$ is described by the creation operator $\hb^\dag(\omega)$ and annihilation operator $\hb(\omega)$ satisfying the commutation relation  $[\hb(\omega),\hb^\dag(\omega')]=\delta(\omega-\omega')$. Assuming that the rate $\kappa$ describing the coupling of the cavity mode to the environment is constant near $\omega_c$, we introduce the photon Hamiltonian~{\color{black}\cite{Gardiner_Qnoise:2004}}
\begin{eqnarray}
\label{H-ph-def}
\hat H_\texttt{ph}&=&
\hbar\int d\omega \omega\hb^\dag(\omega)\hb(\omega)
\\\nonumber 
&-& i\hbar\sqrt{\kappa}\int d\omega\left(\ha^\dag\hb(\omega)-\hb^\dag(\omega)\ha\right),
\end{eqnarray}
whose terms on the right-hand side describe the free photon continuum and its interaction with the cavity mode, respectively. 

{\color{black} The QEs placed inside the cavity interact exclusively with the cavity mode, giving rise to cavity polaritons. In this case, the generic polariton Hamiltonian, $\hat{H}_\texttt{pl}=\hat{H}_\texttt{pl}(\hat{a},\hat{\sigma})$, depends on the cavity mode operator $\hat{a}$ and a set of material dependent QE operators $\hat\sigma$. Accordingly, the total Hamiltonian can be written as
\begin{eqnarray}
\label{H-pl+ph}
\hat H = \hat H_\texttt{pl}+\hat H_\texttt{ph}.
\end{eqnarray}
We further separate the last term in Eq.~\eqref{H-ph-def} from the remaining terms in Eq.~\eqref{H-pl+ph} and represent this contribution as
\begin{eqnarray}
\label{H-pl+ph-I}
\hat H_\text{I}(t) &=& -i\hbar\sqrt{\kappa} e^{\frac{i}{\hbar}\hat H_\texttt{pl}t+i\int d\omega \omega\hb^\dag(\omega)\hb(\omega)t}
\\\nonumber&\times&
\int d\omega'\left[\ha^\dag\hb(\omega')-\hb^\dag(\omega')\ha\right]
\\\nonumber&\times&
e^{-\frac{i}{\hbar}\hat H_\texttt{pl}t-i\int d\omega \omega\hb^\dag(\omega)\hb(\omega)t}.
\end{eqnarray}

Keeping in mind that the polariton Hamiltonian $\hat H_\texttt{pl}$ commutes with its free photon continuum counterpart entering the exponential, Eq.~\eqref{H-pl+ph-I} squires the following form
\begin{eqnarray}
\label{H-ph-int}
\hat H_\text{I} (t)=
-i\hbar\sqrt{\kappa}\int d\omega\left[\ha^\dag (t)\hb(\omega)e^{-i\omega t}
    -\hb^\dag(\omega)\ha (t)e^{i\omega t}\right],\;\;\;\;
\end{eqnarray}
which reflects the interaction representation of the photon continuum operators. The cavity photon operator follows the Heisenberg time evolution $\hat a(t)=e^{i\hat H_\texttt{pl}t/\hbar}\hat a e^{i\hat H_\texttt{pl}t/\hbar}$. As demonstrated below, this formulation allows one to express the continuum photon scattering amplitudes in terms of the cavity photon Green function without specifying the explicit form of $\hat H_\texttt{pl}$, while also enabling the incorporation of QE dissipative dynamics.}

{\color{black} We define the cavity scattering operator as}
\begin{eqnarray}
\label{scat-sef}
\hat S = \left\langle \hat T \exp\left\{-\frac{i}{\hbar}
        \int\limits_{-\infty}^\infty dt ~\hat H_\texttt{I} (t) \right\} \right\rangle,
\end{eqnarray}
where, we use the interacting Hamiltonian in the form of Eq.~\eqref{H-ph-int}, $\hat T$ is the time-ordering operator, and the angular brackets describe average over the polariton degrees of freedom. 

{\color{black} As detailed in Appendix~\ref{App:S-mtx},} we restrict our analysis to the power series expansion of the scattering operator in Eq.~\eqref{scat-sef} up to the lowest order terms that conserve the scattered photon number. Specifically, this includes the trivial zero-order contribution describing the input photon pair passing through the cavity without any interactions with the polaritons, as well as the linear in the coupling rate $\kappa$ contribution
\begin{eqnarray}
\label{scat-1}
\hat S^{(1)} &=& -\kappa \int d\omega_1 \int d\omega_2\int dt_1 \int dt_2 \theta(t_1-t_2)
\\\nonumber&\times&
\left\langle \ha(t_1)\ha^\dag(t_2) \right\rangle e^{i\omega_1t_1-i\omega_2t_2} 
:\hb^\dag(\omega_1)\hb(\omega_2): 
+h.c.,
\end{eqnarray}
describing the single-photon scattering events. The second order terms in the coupling rate $\kappa$,
\begin{eqnarray}
\label{scat-2}
\hat S^{(2)} &=& \kappa^2 \int d\omega_1\dots\int d\omega_4\int dt_1\dots\int dt_4 
\\\nonumber &\times& 
\theta(t_1-t_2) \theta(t_2-t_3)\theta(t_3-t_4)
\\\nonumber&\times&\left\{
\left\langle \ha(t_1)\ha(t_2)\ha^\dag(t_3)\ha^\dag(t_4) \right\rangle 
e^{i\omega_1t_1+i\omega_2t_2-i\omega_3t_3-i\omega_4t_4} 
\right.\\\nonumber &\times&\left.
:\hb^\dag(\omega_1)\hb^\dag(\omega_2)\hb(\omega_3)\hb(\omega_4): 
\right.\\\nonumber&+&\left.
\left\langle \ha(t_1)\ha^\dag(t_2)\ha(t_3)\ha^\dag(t_4) \right\rangle 
e^{i\omega_1t_1-i\omega_2t_2+i\omega_3t_3-i\omega_4t_4} 
\right.\\\nonumber &\times&\left.
:\hb^\dag(\omega_1)\hb(\omega_2)\hb^\dag(\omega_3)\hb(\omega_4):
\right.\\\nonumber&+&\left.
\left\langle \ha^\dag(t_1)\ha(t_2)\ha(t_3)\ha^\dag(t_4) \right\rangle 
e^{-i\omega_1t_1+i\omega_2t_2+i\omega_3t_3-i\omega_4t_4} 
\right.\\\nonumber &\times&\left.
:\hb(\omega_1)\hb^\dag(\omega_2)\hb^\dag(\omega_3)\hb(\omega_4):
\right.\\\nonumber &+&\left. h.c.\right\},
\end{eqnarray}
stand for the biphoton scattering events. In these expressions, $:\hat{O}:$ denotes the operators' normal ordering, and all integral limits range from minus to plus infinity, although the Heaviside step functions, $\theta(t)$, ensure the time-ordering. 

Equation~\eqref{scat-1} describes the process of single photon scattering by cavity polaritons. An incident photon of mode $\omega_1$ gets annihilated, creating a cavity mode at time $t_1$. The polariton Green function 
\begin{eqnarray}
\label{G2t-1}
G(t_1-t_2) &=& -i\theta(t_1 - t_2) \left\langle \ha(t_1) \ha^\dag(t_2) \right\rangle,
\end{eqnarray}
subsequently propagates this mode. At time $t_1$, the cavity mode is annihilated, resulting in the outgoing photon mode of frequency $\omega_2$. Although we propagate the Green function defined by Eq.~\eqref{G2t-1} over the time interval $t_1 - t_2$, it is a two-time correlation function~\cite{Zubarev1960}. This dependence becomes evident at $t_1 = t_2 $, via the initial condition defined by the same-time correlation function $\left\langle \ha(t_2) \ha^\dag(t_2) \right\rangle$. While the same-time correlation function may vary with time, we assume that the cavity is in a steady state (time-independent) population  $\bar n+1\equiv\left\langle \ha\ha^\dag \right\rangle$. This assumption simplifies the subsequent analysis by focusing only on coherent fluctuation propagation interacting with the scattering photons and makes the Green function dependent on only the time difference $t_1-t_2$. The steady-state initial conditions for the Green functions will be explicitly accounted for in Sec.~\ref{Sec:TCJSA}.

Given, the steady-state initial condition, we evaluate the time integrals in Eq.~\eqref{scat-1}, and simplify the single-photon scattering operator to the following form
\begin{eqnarray}
\label{scatGF-1}
\hat{S}^{(1)} &=& -2\pi i \kappa \int d\omega \, G_{}(\omega) :\hb^\dag(\omega) \hb(\omega): + h.c.,
\end{eqnarray}
containing the frequency domain representation of single-polariton Green function
\begin{eqnarray}
\label{GF-1}
G(\omega) &=& \int d\tau \, \theta(\tau) G(\tau) e^{i\omega \tau}.
\end{eqnarray}
The Feynman diagram representing this coherent Rayleigh scattering process is shown in Fig.~\ref{Fig:FD1}~(a).

Partitioning the time-ordered four-point correlation functions in Eq.~\eqref{scat-2} into the products of the time-ordered two-point correlation functions using their semi-group propagation properties~\cite{Gardiner_Qnoise:2004}, reveals two distinct contributions to the biphoton scattering processes. We list them as separate terms of the two-photon scatting operator 
\begin{eqnarray}
\label{scatGF-2}
\hat S^{(2)}=\hat S^{(2)}_\text{c} + \hat S^{(2)}_\text{r}. 
\end{eqnarray}
The first term on the right-hand side describes two-photon coherent Rayleigh scattering events. This contribution derives from the second and third terms in Eq.~\eqref{scat-2}, specifically from the polariton steady-state propagation during the $t_2-t_3$ time interval. After integrating over the time variables, these terms take the form illustrated by the Feynman diagrams in Fig.~\ref{Fig:FD1} (b) and (c). We utilize the Sokhotski–Plemelj formula $(\omega+i0)^{-1}=-i\pi\delta(\omega)+{\cal P}\omega^{-1}$ applied to the steady-state propagator, where $\cal P$ denotes the principal value of the singular integral. The delta-function term yields precisely the coherent biphoton scattering operator   
\begin{eqnarray}
\label{scatGF-20}
\hat S_\text{c}^{(2)} &=& 2\pi^2\kappa^2 \int d\omega_1\int d\omega_2 
\\\nonumber&\times&\left\{-G_{}(\omega_1)G_{}(\omega_2)+G_{}^*(\omega_1)G_{}(\omega_2)\right\}
\\\nonumber &\times&
:\hb^\dag(\omega_1)\hb(\omega_1)\hb^\dag(\omega_2)\hb(\omega_2):
+ h.c..
\end{eqnarray}

The second contribution on the right-hand side of Eq.~\eqref{scatGF-2} denotes the terms that describe spectral energy redistribution. Following the structure of Eq.~\eqref{scat-2}, there are three corresponding terms (and their hermitian conjugates) contributing to the redistribution processes 
\begin{eqnarray}
\label{scatGF-2ir}
\hat S_\text{r}^{(2)} &=& 2\pi\kappa^2 \int d\omega_1\dots\int d\omega_4 
\\\nonumber&~&\hspace{-1.cm}\times\left\{ 
G_{}(\omega_1){\cal G}(\omega_1+\omega_2)G_{}(\omega_4)
\delta(\omega_1+\omega_2-\omega_3-\omega_4)
\right.\\\nonumber&~&\hspace{-1.cm}\times\left.
:\hb^\dag(\omega_1)\hb^\dag(\omega_2)\hb(\omega_3)\hb(\omega_4): 
\right.\\\nonumber&~&\hspace{-1.cm}-\left.
    \left[
        i{\cal P}\frac{G_{}(\omega_1)G_{}(\omega_4)}{\omega_1-\omega_2}
        +G_{}(\omega_1)\bar{\cal G}(\omega_1-\omega_2)G_{}(\omega_4)
    \right]
\right.\\\nonumber&~&\hspace{-1.cm}\times\left.
\delta(\omega_1-\omega_2+\omega_3-\omega_4)
:\hb^\dag(\omega_1)\hb(\omega_2)\hb^\dag(\omega_3)\hb(\omega_4):
\right.\\\nonumber&~&\hspace{-1.cm}+\left.
    \left[
        i{\cal P}\frac{G_{}^*(\omega_1)G_{}(\omega_4)}{\omega_2-\omega_1}
       +G^*_{}(\omega_1)\bar{\cal G}(\omega_2-\omega_1)G_{}(\omega_4)
    \right]
\right.\\\nonumber&~&\hspace{-1.cm}\times\left.
\delta(-\omega_1+\omega_2+\omega_3-\omega_4) 
:\hb(\omega_1)\hb^\dag(\omega_2)\hb^\dag(\omega_3)\hb(\omega_4):
\right.\\\nonumber&~&\hspace{-1.cm}+\left. h.c.\right\},
\end{eqnarray}
represented by diagrams (d)--(f) shown in Fig.~\ref{Fig:FD1}.

The first term in Eq.~\eqref{scatGF-2ir}, represented by diagram (d), accounts for the propagation of bipolariton coherence after interaction with two incident photons. The bipolariton Green function $ {\cal G}(\omega) $ describes this event. In the simplest case, where there is no polariton-polariton interaction, it can be expressed as 
\begin{eqnarray}
\label{G2p-def}
{\cal G}(\omega) &=& 2\pi \int d\omega' G_{}(\omega - \omega') G_{}(\omega').
\end{eqnarray}
The second and third terms in Eq.~\eqref{scatGF-2ir} have a similar two-component structure: the first contribution to each term comes from diagrams (b) and (c), specifically the principal value $\cal P$ term of the steady-state propagator. The second contributions are due to the excited state polariton-polariton coherences entering diagrams (e) and (f). The associated excited state polariton-polariton Green function, in the non-interacting polariton approximation, is 
\begin{eqnarray}
\label{G2pp-1}    
\bar{\cal G}(\omega) &=& 2\pi \int d\omega' G_{}(\omega + \omega') G^*_{}(\omega').
\end{eqnarray}

\begin{figure}[t]
\centering
\includegraphics[width=0.45\textwidth]{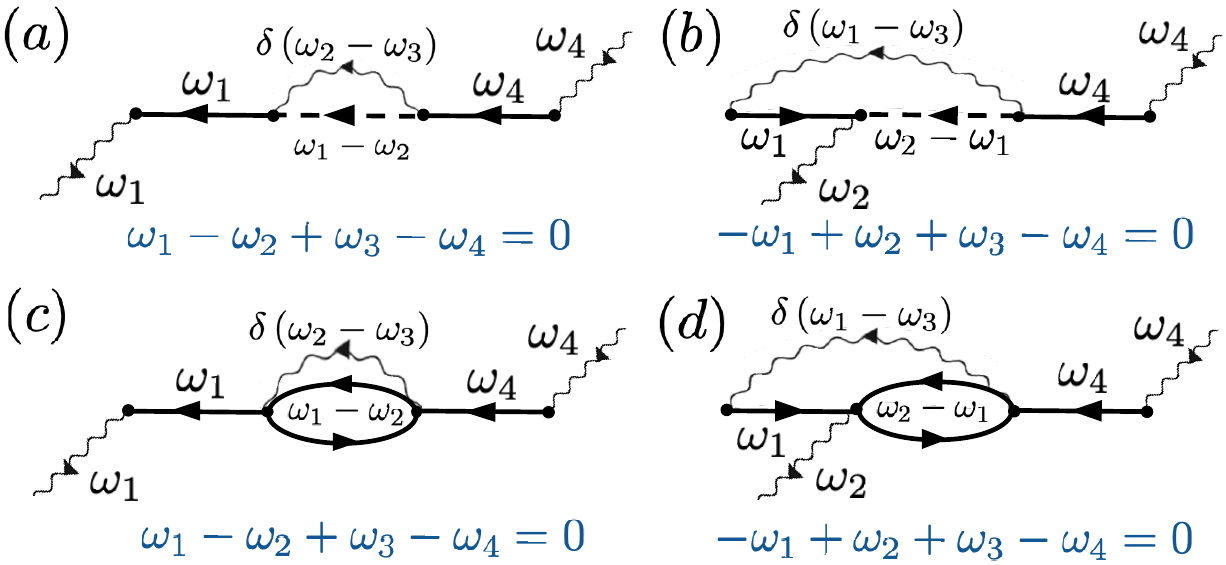}
\caption{Feynman diagrams describing the contributions to the self-energy of the polariton Green function. The correspondence rules are the same as shown in Fig.~\eqref{Fig:FD1}. Here, the connected photon line corresponds to the $\delta$-function, as explicitly indicated in each panel.}
\label{Fig:FD1-corr}
\end{figure}

The normally ordered photon continuum operators, as described in Eqs.~\eqref{scat-1} and \eqref{scat-2}, fully determine the scattering amplitudes presented in Eqs.~\eqref{scatGF-1}, \eqref{scatGF-20}, and \eqref{scatGF-2ir}, as well as in the Feynman diagrams (a) through (f) in Fig.~\ref{Fig:FD1}. Furthermore, we analyzed additional terms that emerge from the permutations (i.e., contractions) of the photon operators as they transition from their initial time-ordered sequences to a normal-ordered state.  {\color{black} These terms give rise to the self-energy in the single-polariton Green function formally derived in Appendix~\ref{App:corr}. The self-energy  accounts for the scattering processes represented by the Feynman diagrams in Fig.~\ref{Fig:FD1-corr}. The physical interpretation of these self-energy contributions is as follows. The Feynman diagrams in Fig.~\ref{Fig:FD1-corr}(a) and (b) contain, connected photon-continuum and cavity steady-state propagators, which explicitly account for the contribution of cavity-mode decay to single-polariton dephasing. The remaining diagrams in Fig.~\ref{Fig:FD1-corr} (c) and (d) provide additional corrections to the polariton dephasing arising from scattering processes involving polariton scattering by another virtual polariton state enabled by the cavity steady-state photon population.} These dephasing rate will be explicitly included in the Green function in Sec.~\ref{Sec:TCJSA}.

\section{Evolution of Biphoton JSA}
\label{Sec:JSA-scat}

Now, we consider an incident biphoton wavepacket 
\begin{eqnarray}
\label{biph-in-t}
|\Psi_\text{in}\rangle &=&\iint d\omega_s d\omega_i {\cal F}_\text{in}(\omega_s,\omega_i)
\hb^\dag(\omega_s)\hb^\dag(\omega_i)|0\rangle,
\end{eqnarray}
prepared via the SPDC process (Fig.~\ref{Fig:Schm}). It is characterized by the JSA ${\cal F}_\text{in}(\omega_s,\omega_i)$ depending on the signal $\omega_s$ and idler $\omega_i$ mode frequencies. The ket $|0\rangle$ represents the vacuum state of photon continuum. 

The output state, resulting from passing this wavepacket through the cavity, is also a biphoton wavepacket obtained by applying the scattering operator to the input state
\begin{eqnarray}
\label{biph-out-t}
|\Psi_\text{out}\rangle &=&\left\{1+\hat S^{(1)}+\hat S^{(2)}\right\}|\Psi_\text{in}\rangle.
\end{eqnarray}
We utilize the expression for  $\hat S^{(1)}$ provided in Eq.~\eqref{scatGF-1} along with  $\hat S^{(2)}=\hat S^{(2)}_\text{c} + \hat S^{(2)}_\text{r}$ specified in Eqs.~\eqref{scatGF-20} and \eqref{scatGF-2ir}, to substitute into Eq.~\eqref{biph-out-t}. After rearranging the photon continuum operators into normal order, we derive the following expression for the output photon wavepacket
\begin{eqnarray}
\label{biph-out-S}
&~&\hspace{-1cm}|\Psi_\text{out}\rangle =\iint d\omega_s d\omega_i 
    \iint  d\omega_s' d\omega_i'~{\cal S}(\omega_s,\omega_i;\omega_{s}',\omega_{i}')
\\\nonumber&~&\times
    {\cal F}_\text{in}(\omega_{s}',\omega_{i}')
    \hb^\dag(\omega_s)\hb^\dag(\omega_i)|0\rangle.
\end{eqnarray}
Here, ${\cal S}(\omega_s,\omega_i;\omega_{s}',\omega_{i}')$ denotes the biphoton scattering matrix.

Following the structure of the photon scattering operator, the scattering matrix comprises two amplitudes
\begin{eqnarray}
\label{S2ph-terms}
{\cal S}(\omega_s,\omega_i;\omega_{s}',\omega_{i}') &=&
    {\cal S}_\text{c}(\omega_s,\omega_i;\omega_{s}',\omega_{i}')
\\\nonumber &+&        
        {\cal S}_\text{r}(\omega_s,\omega_i;\omega_{s}',\omega_{i}').
\end{eqnarray}
The first scattering amplitude describes coherent Rayleigh processes, which combines single-photon scattering in either the signal or idler channels and simultaneous two-photon scattering in each channel. The expression for this coherent scattering amplitude is 
\begin{eqnarray}
\label{S2ph-c}
{\cal S}_\text{c}(\omega_s,\omega_i;\omega_{s}',\omega_{i}') &=& 
\delta(\omega_s-\omega_{s}')\delta(\omega_i-\omega_{i}')
\\\nonumber &~&\hspace{-1.3cm}\times
\left\{1-2\pi i\kappa\left[G_{}(\omega_{s})-G_{}^*(\omega_{s})\right]\right\}
\\\nonumber &~&\hspace{-1.2cm}\times
\left\{1-2\pi i\kappa\left[G_{}(\omega_{i})-G_{}^*(\omega_{i})\right]\right\}.
\end{eqnarray}
The second term on the right-hand side of Eq.~\eqref{S2ph-terms} is biphoton scattering amplitude that describes the processes of incoherent redistribution. Its explicit representation is
\begin{eqnarray}
\label{S2ph-r}
{\cal S}_\text{r}(\omega_s,\omega_i;\omega_{s}',\omega_{i}') &=& 
    4\pi\kappa^2\delta(\omega_s+\omega_i-\omega_{s}'-\omega_{i}')
\\\nonumber&~&\hspace{-2.5cm}\times\left\{    
   G_{}(\omega_s) {\cal G}(\omega_s+\omega_i)
    \left[G_{}(\omega_{s}')+G_{}(\omega_{i}')\right]
\right.\\\nonumber&~&\hspace{-2.5cm}+\left.
    \left[G_{}^*(\omega_{i}')-G(\omega_s)\right]
        \left[{\cal P}\frac{i}{\omega_s-\omega_{i}'}+\bar{\cal G}(\omega_s-\omega_{i}')\right]G_{}(\omega_{s}')
\right.\\\nonumber&~&\hspace{-2.5cm}+\left.
       \left[ G_{}^*(\omega_{s}')-G(\omega_s) \right]
       \left[ {\cal P} \frac{i}{\omega_s-\omega_{s}'}+\bar{\cal G}(\omega_s-\omega_{s}')\right]G_{}(\omega_{i}')
\right.\\\nonumber&~&\hspace{-2.5cm}\left.
\right\}+c.c..
\end{eqnarray}

Similar to Eq.~\eqref{biph-in-t}, we can define the scattered biphoton state
\begin{eqnarray}
\label{biph-out-ww}
|\Psi_\text{out}\rangle =
\iint d\omega_{s} d\omega_{i} {\cal F}_\text{out}
(\omega_s,\omega_i)\hb^\dag(\omega_s)\hb^\dag(\omega_i)|0\rangle,
\end{eqnarray}
designated by the output JSA, ${\cal F}_\text{out}(\omega_s,\omega_i)$. Further comparing this expression with the expression for the output biphoton state given by Eq.~\eqref{biph-out-S}, we identify the following relationship between the input and output JSAs
\begin{eqnarray}
\label{JSA-out-def}
&~&\hspace{-1cm}
{\cal F}_\text{out}(\omega_{s},\omega_{i})=
\iint d\omega_{s}' d\omega_{i}'
\\\nonumber&\times&
 {\cal S}(\omega_s,\omega_i;\omega_{s}',\omega_{i}')
{\cal F}_\text{in}(\omega_{s}',\omega_{i}').
\end{eqnarray}

Finally, utilizing the expressions for scattering amplitudes provided by Eqs.~\eqref{S2ph-terms}--\eqref{S2ph-r}, we substitute them into Eq.~\eqref{JSA-out-def} and integrate over all delta-functions describing the energy conservation. As a result, we represent the output JSA as sum of two components
\begin{eqnarray}
\label{JSA-out-part}
&~&\hspace{-1cm}
{\cal F}_\text{out}(\omega_{s},\omega_{i})=
{\cal F}_\text{c}(\omega_{s},\omega_{i})+{\cal F}_\text{r}(\omega_{s},\omega_{i}).
\end{eqnarray}
reflecting the effect of coherent Rayleigh scattering
\begin{eqnarray}
\label{JSA-c}
&~&\hspace{-1cm}
{\cal F}_\text{c}(\omega_{s},\omega_{i})=
    \left\{1-2\pi i\kappa\left[G_{}(\omega_{s})-G_{}^*(\omega_{s})\right]\right\}
\\\nonumber &~&\hspace{0.cm}\times
    \left\{1-2\pi i\kappa\left[G_{}(\omega_{i})-G_{}^*(\omega_{i})\right]\right\}
    {\cal F}_\text{in}(\omega_{s},\omega_{i}).
\end{eqnarray}
and the incoherent spectral redistribution
\begin{eqnarray}
\label{JSA-r}
{\cal F}_\text{r}(\omega_{s},\omega_{i}) &=& 
    4\pi\kappa^2\int d\omega'
\\\nonumber&~&\hspace{-2.0cm}\times\left\{    
   G_{}(\omega_s) {\cal G}(\omega_s+\omega_i) G_{}(\omega')
\right.\\\nonumber&~&\hspace{-2.0cm}\times \left.  
    \left[{\cal F}_\text{in}(\omega',\omega_s+\omega_i-\omega')
    +{\cal F}_\text{in}(\omega_s+\omega_i-\omega',\omega')\right]
\right.\\\nonumber&~&\hspace{-2.0cm}+\left.
       \left[ G_{}^*(\omega_{s}-\omega')-G(\omega_s) \right]
       \left[ {\cal P} \frac{i}{\omega'}+\bar{\cal G}(\omega')\right]G_{}(\omega_{i}+\omega')
\right.\\\nonumber&~&\hspace{-2.0cm}\times \left.        
       \left[{\cal F}_\text{in}(\omega_s-\omega',\omega_i+\omega')
       +{\cal F}_\text{in}(\omega_i+\omega',\omega_s-\omega')\right]
%
\right\}+c.c..
\end{eqnarray}
Eqs.~\eqref{JSA-out-part} -- \eqref{JSA-r} represent our main theory result, providing a representation of the JSA evolution through polariton Green functions that account for both coherent and incoherent biphoton scattering pathways. To proceed with further analysis, we need the explicit representation of the Green functions, which is defined by a specific model for the interacting cavity mode and QEs, as is done in the next section.  

The form of Eq.~\eqref{JSA-r} indicates that the output JSA is not symmetric with respect to the permutation of the signal and idler channels in the argument. However, the input JSA in Eq.~\eqref{JSA-r} naturally becomes symmetric due to the indistinguishability of signal and idler photons. For this reason, we can also interchange the channels in the definition of the output JSA. Consequently, we can  express the outgoing biphoton JSA as a linear combination $\alpha {\cal F}_\text{out}(\omega_s, \omega_i) + \beta {\cal F}_\text{out}(\omega_i, \omega_s)$, for  $\{\alpha,\beta\} \in \mathbb{C}$ with $\alpha + \beta = 1$. The parameters can be assessed by calculating the photon transmission functions of the TWINS interferometers, as well as the response of the photon detectors illustrated in Fig.~\ref{Fig:Schm}. Since this paper focuses on developing the theory for biphoton scattering within the cavity, we will not take into account the specific effects of the instruments. Instead, we will further investigate the symmetrized form of the output JSA~\footnote{For some calculations (such as the contour integration leading to Eq. \eqref{JSA-TCr} ), it is convenient to use an unsymmetrized JSA. However, to make fair comparisons between input and output JSA and their associated entanglement, it is important to compare the symmetrized versions only, as the unsymmetrized versions can be changed arbitrarily.} 
\begin{equation}
\label{Sym-JSA-def}
    {\cal F}_\text{out}^{sym}(\omega_s, \omega_i) = \frac{1}{2}\left[{\cal F}_\text{out}(\omega_s,\omega_i) + {\cal F}_\text{out}(\omega_i, \omega_s)\right].
\end{equation}

Finally, the entanglement of the input/output JSA can be quantified by performing its Schmidt decomposition,
\begin{equation}
\label{schmidt_decomp}
\mathcal{F}(\omega_s, \omega_i) = \sum_j r_j g_j(\omega_s)h_j(\omega_i),
\end{equation}
which is numerically implemented by discretizing the JSA arguments and subsequently performing a matrix singular value decomposition (SVD). In this decomposition, $\{g_j\}$ and $\{h_j\}$ are the SVD basis vectors, and the $\{r_j\}$ is a set of the Schmidt coefficients. The entanglement entropy of the JSA is then given by
\begin{equation}
S(\mathcal{F}) = -\sum_j \tilde r_j^2\ln(\tilde r_j^2),
\end{equation}
where
\begin{equation}
\label{schmidt_coef}
\tilde{r}_j = \frac{r_j}{\sqrt{\sum_j r_j^2}}
\end{equation}
denotes the normalized Schmidt coefficients.

\section{Application to Tavis-Cummings model}
\label{Sec:TCJSA}

For further analysis, we adopt the Tavis-Cummings (TC) model to describe the cavity polaritons. This model considers a cavity containing an ensemble of $N$ identical two-level QEs, each located at a site indexed by $n$. Considering their coupling to the cavity mode within the rotating-wave approximation. {\color{black} Therefore, we identify the polariton Hamiltonian, $\hat H_\texttt{pl}$, entering Eq.~\eqref{H-pl+ph} as the TC Hamiltonian, $\hat H_\texttt{pl}\equiv \hat H_\text{TC}$ where }
\begin{eqnarray}
\label{H-TC-def}
\hspace{-0.5cm} \hat H_\text{TC}&=&\hbar \omega_c \ha^\dag\ha
    +\hbar\omega_o \sum_{n=1}^{N}\hsp_n\hsm_n
\\\nonumber
    &+& \hbar \lambda\sum_{n=1}^{N}\left(\ha^\dag\hsm_n+\hsp_n\ha\right).
\end{eqnarray}
Here, $\omega_o$ and $\omega_c$ are the QE transition frequency and the cavity mode central frequency, respectively. The two-level QEs are described by the Pauli SU(2) operators $\hat\sigma_n^\pm=(\hsx_n\pm i\hsy_n)/2$ and $\hsz_n$, while the cavity mode by the boson creation, $\hat a^\dag$, and annihilation, $\hat a$, operators; $\lambda$ is the QE-cavity mode coupling rate.

The derivation of the polariton Green functions for the TC-model is provided in Appendix~\ref{Appx:TCModel}. For our analysis, we use their representation utilizing sum over poles, $\tilde\omega_\pm$, 
\begin{eqnarray}
\label{Gapls}
&~&G_{}(\omega)=\sum_{\alpha=\pm}\frac{u_\alpha}{2\pi(\omega-\tilde\omega_\alpha)},
\\\label{Gapls-cc}  
&~& G^*_{}(\omega)=\sum_{\alpha=\pm}\frac{u^*_\alpha}{2\pi(\omega-\tilde\omega^*_\alpha)},
\end{eqnarray}
where the polariton expansion coefficients are
\begin{eqnarray}
\label{upm-def}
u_\pm &=&\pm\frac{\left(\tilde\omega_\pm-\tilde\omega_o\right)\left(\bar n+1\right)+\lambda N~\overline{\sigma a}}
    {\tilde\omega_{+}-\tilde\omega_{-}},
\\\label{upm-cc}  
u^*_\pm &=&\pm\frac{\left(\tilde\omega^*_\pm-\tilde\omega^*_o\right)\bar{n}+\lambda N~\overline{\sigma a}^*}    {\tilde\omega^*_{+}-\tilde\omega^*_{-}}.
\end{eqnarray}
These coefficients explicitly account for the initial conditions such as the steady-state population of the cavity mode, denoted as $\bar n$, and the steady-state coherence between the QE and the cavity mode $ \overline{\sigma a}\equiv\sum_{n=1}^N\left\langle \hsm_n\ha^\dag\right\rangle/N$. Typically, this coherence is zero unless the symmetry of the Hamiltonian is spontaneously broken, which then leads to its emergence~\cite{Li_QSciTec:2018,Kirton_AQT:2019,Piryatinski_PRR:2020}.  It is important to note that what we denote as the complex conjugate expansion coefficient, as provided by Eq.~\eqref{upm-cc}, includes the steady-state population factor $\bar n$, in contrast to $\bar n + 1$ in Eq.~\eqref{upm-def}. This difference arises from the initial condition for the complex conjugate Green's function, which is determined by the correlation function $\langle\ha^\dag \ha\rangle \equiv \bar n$. We will use this subtle property in Sec.~\ref{Sec:Res} to obtain Eqs.~\eqref{F_c_vac} and \eqref{F_r_vac} for the case of cavity vacuum, $\bar n =0$.

The poles of the Green function 
\begin{eqnarray}
\label{wpm-def}
\tilde\omega_\pm=\frac{\tilde\omega_c+\tilde\omega_o}{2}
        \pm\sqrt{\frac{\left(\tilde\omega_c-\tilde\omega_o\right)^2}{4}-\lambda^2N\bsz},
\end{eqnarray}
describe the upper-polariton, $\tilde\omega_{+}=\omega_{+}-i\gamma_{+}$, and lower polariton $\tilde\omega_{-}=\omega_{-}-i\gamma_{-}$ complex eigenfrequencies. They depend on the QE steady-state population inversion $\bar\sigma_z=2\sum_{n=1}^N\left\langle \hsp_n\hsm_n\right\rangle/N-1$ and complex frequencies of the cavity mode, $\tilde\omega_c = \omega_c - i\gamma_c$, and QEs, $\tilde\omega_o = \omega_o - i\gamma_o$. The dephasing rate of the QE, $\gamma_o$, is a phenomenological parameter. {\color{black} Here, we use the cavity dephasing rate, $\gamma_c=\kappa\left[1/2+\bar{n}\left(\bar{n}+1\right)\right]$,  derived in Appendix~\ref{App:corr}  as the frequency independent part of the self-energy term arising from the coupling between the cavity mode and photon continuum. The first term is due to the Feynman diagram (a) in Fig.~\ref{Fig:FD1-corr} and the second term in this expression explicitly  accounts for polariton scattering processes described by the Feynman diagrams in Fig.~\ref{Fig:FD1-corr} (c)-(d).} 

The bipolariton and excited state polariton coherence Green functions are derived using Eqs.~\eqref{G2p-def} and \eqref{G2pp-1}. After evaluating the convolution integrals in these equations with the polariton Green functions provided in Eqs.~\eqref{Gapls} and \eqref{Gapls-cc}, we obtain the following expressions 
\begin{eqnarray}
\label{Gdc-11}
{\cal G}(\omega) &=& -i\sum_{\alpha=\pm}\sum_{\beta=\pm}
        \frac{u_\alpha u_\beta}{\omega-\tilde\omega_\alpha-\tilde\omega_\beta},
\\\label{Gdc-12}
    \bar{\cal G}(\omega) &=&i\sum_{\alpha=\pm}\sum_{\beta=\pm}
        \frac{u_\alpha u^*_\beta}{\omega-\tilde\omega_\alpha+\tilde\omega^*_\beta}.       
\end{eqnarray}
The bipolariton coherence Green function, ${\cal G}_{}(\omega)$, has three poles that represent all possible summations of the single polariton frequencies. The poles of the excited state polariton-polariton coherence Green function, $\bar{\cal G}(\omega)$, contain all possible differences between single polaritons frequencies. 

Given the explicit representation of the single polariton Green functions in Eqs. \eqref{Gapls} -- \eqref{upm-cc}, the calculation of the coherent output JSA can be carried out by substituting these equations into Eq. \eqref{JSA-c}. For the  component, following Eq. \eqref{JSA-r}, a convolution integral must be computed after incorporating Eqs.~\eqref{Gapls}–\eqref{upm-cc}, as well as Eqs.~\eqref{Gdc-11} and \eqref{Gdc-12} for all the relevant Green functions. This integral can be calculated numerically. However, in the case of the TC model, the number of poles in the Green functions is limited, which allows us to evaluate the convolution integral explicitly.

\begin{figure*}[t]
\includegraphics[width=0.85\textwidth]{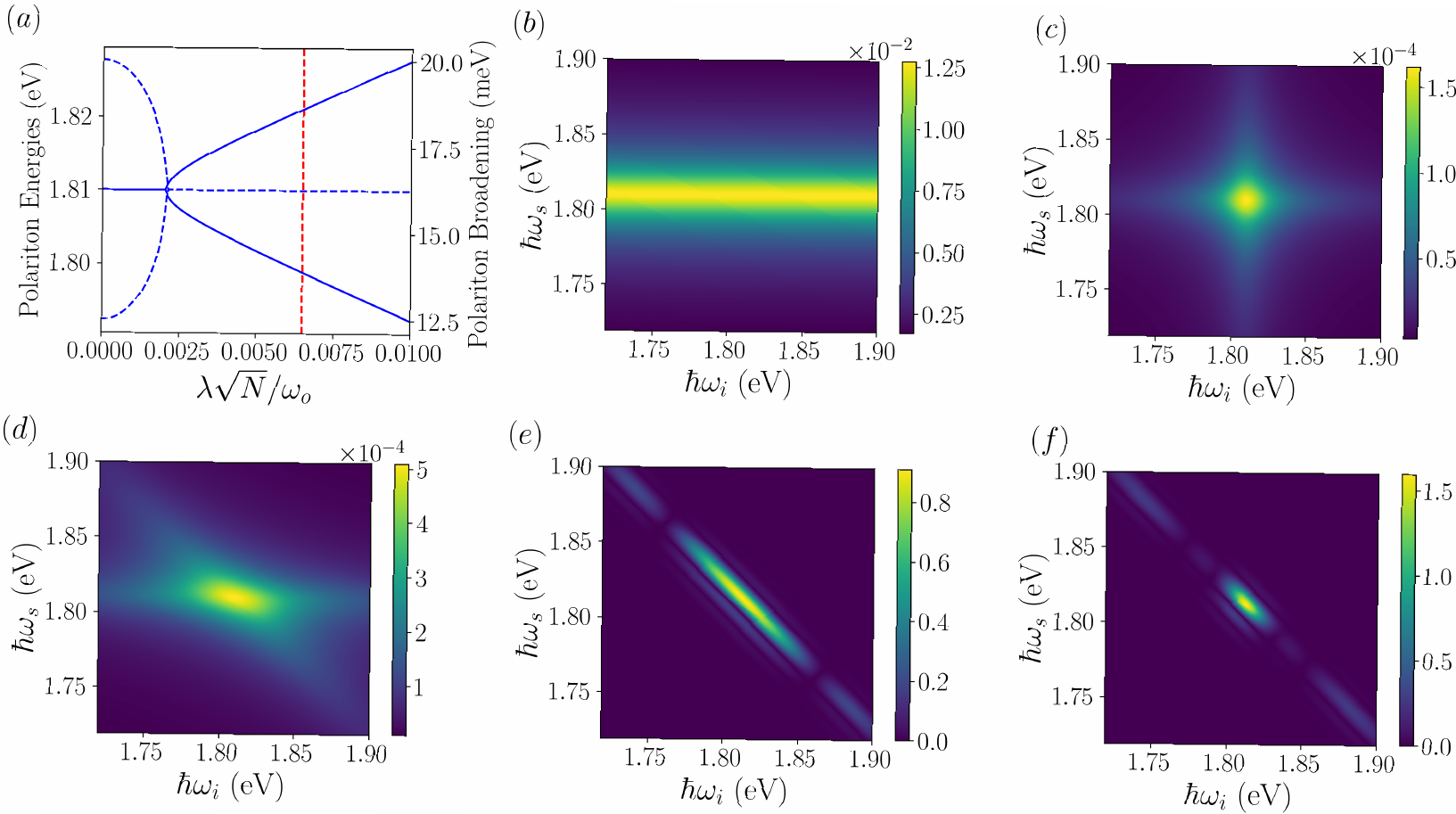}
    \caption{(a) Calculated dispersion of the polariton energy $\hbar\omega_\pm$ (solid blue) and broadening $\hbar\gamma_\pm$ (dashed blue) for the cavity in vacuum state, $\bar{n}=0$ as a function of the  normalized cooperative coupling parameter $\lambda\sqrt{N}/\omega_o$. The vertical dashed red line marks $\lambda\sqrt{N}/\omega_o = 0.0065$. The rest of the panels present calculation results for the empty cavity case, $\lambda\sqrt{N}=\bar{n}=0$: (b)~$|G(\omega_s)|$, (c)~$|G(\omega_s)G(\omega_i)|$, (d)~  $|G(\omega_s)\mathcal{G}(\omega_s + \omega_i)|$, (e) Absolute value of symmetrized input (SPDC) JSA, (f) Absolute value of symmetrized output JSA.
    }
    \label{Fig:empty_cav}
\end{figure*}

For this and subsequent calculations, we adopt the following functional form of the input JSA that describes the SPDC processes in a nonlinear crystal with a side length $L$
\begin{equation}
\label{SPDC_theory_input}
\mathcal{F}_\textrm{in}(\omega_s, \omega_i) \equiv \Gamma(\omega_s + \omega_i)
\text{sinc}\left(\frac{\Delta k_\parallel L}{2}\right)\text{sinc}\left(\frac{\Delta k_\perp L}{2}\right).
\end{equation}
Equation~\eqref{SPDC_theory_input} contains $\Gamma(\omega) = \exp\{-(\omega - \omega_p)^2/4\sigma_p^2\}$, the Gaussian pump pulse envelope function with the central frequency $\omega_p$ and spectral variance $\sigma_p^2$; the product of the sinc$(x)=\sin(x)/x$ functions describe the phase-matching conditions in both directions parallel and perpendicular to the incident pump photons. The associated wavevector mismatches are  $\Delta k_{\parallel} \approx [n(\omega_p) \omega_p - n(\omega_p/2)\left(\omega_s\cos(\theta_1) + \omega_i\cos(\theta_2)\right)]/c$ and $\Delta k_{\perp} = n[\omega_p/2)(\omega_s\sin(\theta_1)-\omega_i\sin(\theta_2)]/c$, with $\theta_1$ and $\theta_2$ being the collection angles, $n(\omega)$ the refraction index of the crystal, and $c$ the speed of light~\cite{Moretti_JCP:2023}.

After labeling the angles so that $\cos(\theta_1) \leq \cos(\theta_2)$ and performing contour integration of the convolution integrals, we obtain the sum-over-polariton-pole expression for the incoherent redistribution JSA, which reads
\begin{widetext}
\begin{eqnarray}
\label{JSA-TCr}
&~&\hspace{-0.5cm}{\cal F}_\textrm{r}(\omega_s, \omega_i) = 
    4\pi^2\kappa^2\left[G^*(\omega_s)G(\omega_i)-G(\omega_s)G(\omega_i)\right]\left[{\cal F}_{\textrm{in}}(\omega_s, \omega_i) - {\cal F}_{\textrm{in}}(\omega_i, \omega_s)\right] 
\\\nonumber &+&
    4\pi\kappa^2\sum_{\alpha = \pm} u_\alpha \left\{
    \left[G(\omega_s) - G^*(\omega_s + \omega_i - \tilde\omega_\alpha)\right] 
    \left[i\bar{\mathcal{G}}(\omega_i - \tilde\omega_\alpha) - \frac{1}{\tilde\omega_\alpha - \omega_i} \right]
    - i G(\omega_s)\mathcal{G}(\omega_s + \omega_i)\right\}{\cal F}_{\textrm{in}}(\omega_s + \omega_i - \tilde\omega_\alpha, \tilde\omega_\alpha) 
\\&+&\nonumber  
    4\pi\kappa^2\sum_{\alpha = \pm} u_\alpha^* G(\omega_s + \omega_i - \tilde\omega_\alpha^*)\left[i \bar{\cal G}(\omega_s - \tilde\omega_\alpha^*) - \frac{1}{\omega_s - \tilde\omega_\alpha^*}\right]\mathcal{F}_{\textrm{in}}(\tilde\omega_\alpha^*, \omega_s + \omega_i - \tilde\omega_\alpha^*) 
\\&-&\nonumber 
    8\pi^2\kappa^2\sum_{\alpha = \pm}\sum_{\beta = \pm} u_\alpha u_\beta^* \left[G(\omega_s) - G^*(\omega_s - \tilde\omega_\alpha + \tilde\omega_\beta^*)\right]G(\tilde\omega_\alpha - \tilde\omega_\beta^* + \omega_i) {\cal F}_{in}(\omega_s - \tilde\omega_\alpha + \tilde\omega_\beta^*, \omega_i + \tilde\omega_\alpha - \tilde\omega_\beta^*) + \nonumber c.c..
    \end{eqnarray}
\end{widetext}
In the complex conjugate part of this expression, the arguments of $\mathcal{F}_{in}$ must be complex conjugated and permuted. For example, $\mathcal{F}_{in}(\omega_s + \omega_i - \tilde\omega_\alpha, \tilde\omega_\alpha) \to \mathcal{F}_{in}(\tilde\omega_\alpha^*, \omega_s + \omega_i - \tilde\omega_\alpha^*)$. Additionally notice that the first term is proportional to ${\cal F}_{\textrm{in}}(\omega_s, \omega_i) - {\cal F}_{\textrm{in}}(\omega_i, \omega_s)$. {\color{black} Accordingly, it vanishes upon symmetrization of the output JSA accounting for the indistinguishability of the outgoing signal and idler photons, and proposed to evaluate the entanglement entropy in Sec.~\ref{Sec:JSA-scat}.}

\section{Results} 
\label{Sec:Res}

\begin{figure*}[t]
\includegraphics[width=0.85\textwidth]{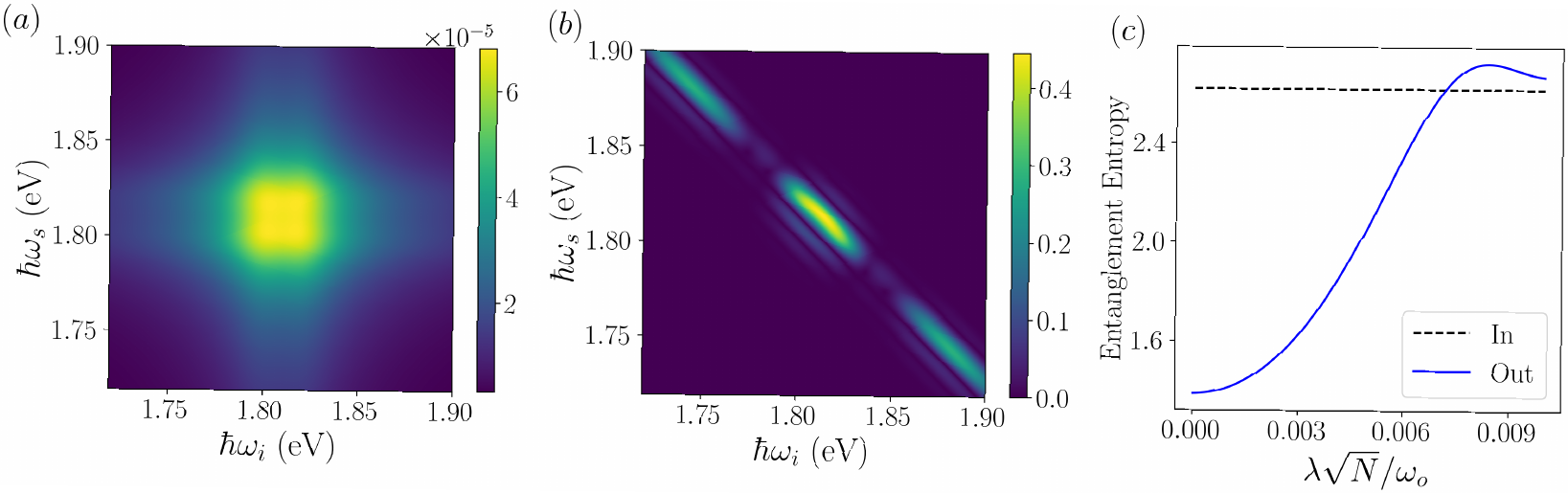}
    \caption{ Calculated (a) $|G(\omega_s)G(\omega_i)|$ and (b) absolute value of symmetrized output JSA, for the case of QEs in a strong coupling regime, $\lambda\sqrt{N}/\omega_o = 0.0065$, [designated by vertical red dashed line in Fig.~\ref{Fig:empty_cav}(a)] with the cavity mode in vacuum state, $\bar{n}=0$.
    (c) Comparison of the entanglement entropy due to the symmetrized output JSA presented in panel~(b) with the entanglement entropy of the symmetrized input JSA presented in Fig.~\ref{Fig:empty_cav}(e) v.s. normalized cooperative coupling parameter $\lambda\sqrt{N}/\omega_o$.
    }
    \label{Fig:StongCoup}
\end{figure*}

To calculate the polariton and bipolariton Green functions introduced in Section~\ref{Sec:TCJSA}, we parameterize the TC model as follows. The transition energy of the QEs and the energy of the cavity mode are set to be identical and equal to $\hbar\omega_o = \hbar\omega_c = 1.81$ eV. The QE dephasing and cavity leakage, energy are set to $\hbar\gamma_o = 20$ meV and $\hbar\kappa = 25$ meV, respectively. Throughout the whole analysis, we consider the QE steady state to be the ground state, defined by $\bar{\sigma_z} = -1$. {\color{black} Below, the steady-state population of cavity photons, $\bar{n}$, is varied over a very small range, $0 \leq \bar n \leq 2$, due to the incoherent pumping of the QEs. Since the number of QEs in the cavity is $N \gg 1$, the corrections to the QE ground-state populations which scale as $\bar n/N$ can be neglected. Given that no QE population inversion is present, $\bar{\sigma}_z < 0$, the system resides in the trivial ground state, $\bar{\sigma}_\pm = \bar{a} = 0$, and therefore no spontaneous coherences between the QEs and the cavity mode can emerge within the TC model~\cite{Li_QSciTec:2018,Kirton_AQT:2019,Piryatinski_PRR:2020,Sukharev_JCP:2021,Piryatinski_Ntech:2023}, allowing us to set $\overline{\sigma a} = 0$}. 

In the following, we examine the impact of biphoton scattering on the symmetrized output JSA [Eq.~\eqref{Sym-JSA-def}] for two different types of input JSAs, considering various values of the cavity steady-state population $\bar{n}$ and the cooperative coupling parameter $\lambda\sqrt{N}$. {\color{black} Although photons associated with the cavity steady-state occupancy leave the cavity and may, in principle, contribute directly to the detected signal, we neglect this contribution in the JSA calculations. The reason is that the entangled biphoton states can be effectively distinguished from this steady-state background through coincidence counting methodology.}

\subsection{Model input JSA}

 The SPDC input JSA is calculated using Eq.~\eqref{SPDC_theory_input}. The pump pulse energy is set to $\hbar\omega_p = 3.62$~eV, and the pulse variance is $\hbar\sigma_p = 10$~meV. The length of the nonlinear crystal is   fixed at $L = 0.1$~mm, the refractive index at $n(\omega) = 1$, and the collection angles at $\theta_1 = \theta_2 = 3.5^{\circ}$. The absolute value of this JSA is shown in Fig.~\ref{Fig:empty_cav}(e), where its width along the diagonal direction ($\omega_s=\omega_i$) is determined by the nonlinear crystal phase-matching conditions, while its elongation in the counter-diagonal direction ($\omega_s+\omega_i=\omega_p$) is governed by the pump pulse variance $\sigma_p$.

To start, we will examine the case of a cavity in the vacuum state $\bar n = 0$. This reduces Eqs. \eqref{JSA-c} and \eqref{JSA-TCr} describing the output JSA components to the form
\begin{eqnarray}
\label{F_c_vac}
{\cal F}_\text{c}(\omega_{s},\omega_{i})&=& [1-2\pi i\kappa G(\omega_{s})][1-2\pi i\kappa G(\omega_{i})]
\\\nonumber &\times&
    {\cal F}_\text{in}(\omega_{s},\omega_{i}),
\\\label{F_r_vac}
    {\cal F}_{\textrm{r}}(\omega_s, \omega_i) &=& -4\pi\kappa^2 G(\omega_s)\sum_{\alpha = \pm} u_\alpha \left[\frac{1}{\tilde\omega_\alpha - \omega_i}
\right.\\\nonumber &+& \left.    
    i\mathcal{G}(\omega_s + \omega_i) \right]  
    {\cal F}_{in}(\omega_s + \omega_i - \tilde\omega_\alpha, \tilde \omega_\alpha),
\end{eqnarray}
respectively. According to Eq.~\eqref{F_r_vac}, the incoherent redistribution occurs solely due to the interplay between the principal value of the ground state propagator, which appears in the Feynman diagram (b), and the bipolariton coherence present in the Feynman diagram (d), both shown in of Fig.~\ref{Fig:FD1}. The term in Eq.~\eqref{JSA-TCr} that is proportional to ${\cal F}_{\textrm{in}}(\omega_s, \omega_i) - {\cal F}_{\textrm{in}}(\omega_i, \omega_s)$ has been omitted in Eq.~\eqref{F_r_vac} because it does not contribute to the symmetrized output JSA.

Figure~\ref{Fig:empty_cav}(a) presents the dispersion of polariton energy and broadening parameter calculated according to Eq.~\eqref{wpm-def} as a function of the TC cooperative coupling parameter varying in the range of $0 < \lambda\sqrt{N} < 0.01\omega_o$. For $\lambda\sqrt{N} \leq 0.0021\omega_o$, the cavity is in the weak coupling regime, characterized by the bare energies of the cavity mode and QEs, along with their broadening parameters. Once $\lambda\sqrt{N} > 0.0021\omega_o$, the cavity transitions into the strong coupling regime, which is marked by the Rabi splitting that can be seen in the plot. In this regime, the broadening parameter for the polaritons becomes $\hbar\gamma_{+} = \hbar\gamma_{-} = 16.25$ meV. 

The simplest case of an empty cavity, $\lambda\sqrt{N} = \bar{n} = 0$, provides a starting point for understanding the relationship between the input and output JSAs arising from biphoton scattering. In this trivial case the single polariton state described by the poles of the Green function $G(\omega)$ [Eq.~\eqref{Gapls}] is the broadened cavity mode resonance at $\omega_c-i\kappa/2$. Furthermore, the pole of the bipolariton coherence Green function, ${\cal G}(\omega)$ [Eq.~\eqref{Gdc-11}] is merely the double excitation of the cavity mode. According to Eqs.~\eqref{F_c_vac} and \eqref{F_r_vac}, the resonances associated with the Green function combinations illustrated in Figs.~\ref{Fig:empty_cav}(b)–(d), after convolution with the input JSA shown in Fig.~\ref{Fig:empty_cav}(e), give rise to a complex Fano-type resonant response. This response defines the output JSA presented in Fig.~\ref{Fig:empty_cav}(f). 

The cavity response can be interpreted as a spectral filtering effect. The maximum of the output JSA [Fig.~\ref{Fig:empty_cav}(f)] is centered around the cavity resonance at $\hbar\omega_s = \hbar\omega_i = 1.81~\mathrm{eV}$. This point corresponds to the intersection between the input JSA and all resonances shown in panels (b)–(d). The resonance structures displayed in panels (b) and (c) of Fig.~\ref{Fig:empty_cav} are characteristic of the coherent JSA component described by Eq.~\eqref{F_c_vac} and the first (bracketed) contribution to the incoherent JSA given by Eq.~\eqref{F_r_vac}. The only resonance that extends along the counter-diagonal direction, i.e., aligned with the input JSA, appears in panel (d). This feature originates from the pole of the bipolariton coherence Green function, $\omega_s + \omega_i = 2\omega_c - i\kappa$, which enters the incoherent redistribution component of the output JSA [Eq.~\eqref{F_r_vac}]. Since this contribution scales as $\kappa^2$, isolating it from an experimentally  measured JSA would require subtracting the coherent term [Eq.~\eqref{F_c_vac}] contribution.

Next, we examine the effect of strong coupling between the cavity in vacuum state, $\bar{n}=0$, and the QEs, facilitated by the coupling parameter $\lambda\sqrt{N}=0.0065\omega_o$. This regime is indicated by the red dashed line in the polariton dispersion [Fig.~\ref{Fig:empty_cav}(a)]. Compared to the empty-cavity case, we now observe two Rabi-split polariton resonances at $\omega_{+}=1.821$~eV and $\omega_{-}=1.799$~eV, along with an increased polariton broadening from $\hbar\kappa/2=12.5$~meV to $\hbar\gamma_{+} = \hbar\gamma_{-} = 16.25$~meV. The calculated pole structure of the single-polariton Green function product, $|G(\omega_s)G(\omega_i)|$, is presented in Fig.~\ref{Fig:StongCoup}(a). Compared to Fig.~\ref{Fig:empty_cav}(c), a clear signature of Rabi splitting and increased linewidth broadening is observed. Accordingly, the output JSA shown in Fig.~\ref{Fig:StongCoup}(b) exhibits enhanced features in the vicinity of the polariton resonances, relative to the empty-cavity case in Fig.~\ref{Fig:empty_cav}(f). This observation also imposes a limitation on the probing of the cavity resonances. Specifically, the Rabi splitting should not exceed the width of the input JSA in the counter-diagonal direction determined by the SPDC pump pulse variance, i.e., $\omega_{+}-\omega_{-} < \sigma_p$.

In the adopted strong-coupling regime, the poles of the bipolariton Green function give rise to three resonances, $\omega_s+\omega_i=2(\omega_\pm-i\gamma_\pm)$ and $\omega_s+\omega_i=2(\omega_{+}+\omega_{-}-i\gamma_{+}-i\gamma_{-})$, which extend along the counter-diagonal direction. However, they cannot be resolved in the output JSA [Fig.~\ref{Fig:StongCoup}(b)]. Similar to the empty-cavity case, the contribution of these resonances can be isolated by subtracting the coherent component from the total output JSA.  

Figure~\ref{Fig:StongCoup}(c) shows the input and output JSA entanglement entropy, calculated according to Eqs.~\eqref{schmidt_decomp}--\eqref{schmidt_coef}, as a function of the coupling strength $\lambda\sqrt{N}$. In Appendix~\ref{App:Schidt}, we present arguments demonstrating that merely the broadening of the output JSA, observed in Fig.~\ref{Fig:StongCoup}(b) relative to Fig.~\ref{Fig:empty_cav}(f), leads to an increase in the corresponding entanglement entropy. Consequently, as the polariton states evolve from the empty-cavity to the strong-coupling regime within the range $0 \leq \lambda\sqrt{N} \leq 0.008\omega_o$, the entropy in Fig.~\ref{Fig:StongCoup}(c) increases. Beyond $\lambda\sqrt{N}/\omega_o = 0.008$, the entanglement saturates due to the suppression of the output JSA because the polariton splitting exceeds the pump pulse variance.

\begin{figure}[t]
\includegraphics[width=0.45\textwidth]{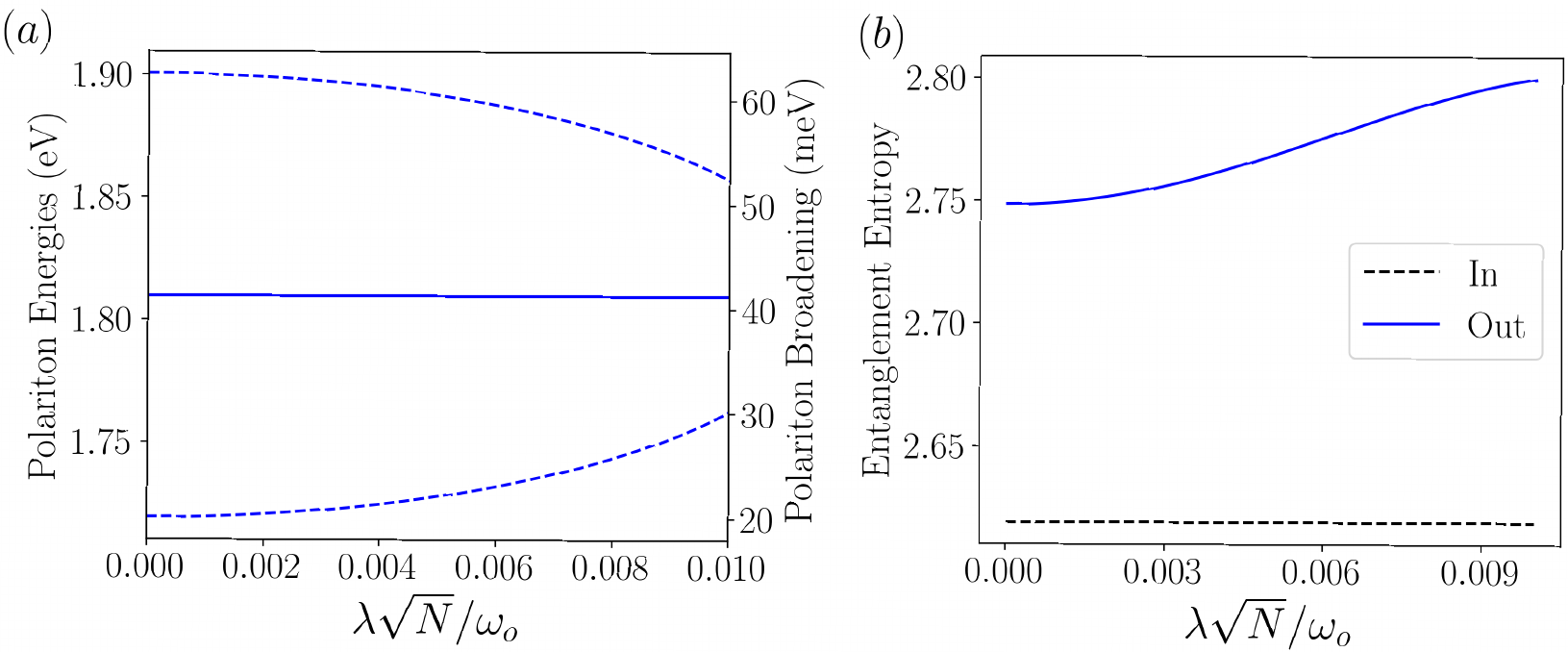}
    \caption{ Calculation results for the case of cavity containing a single photon, $\bar{n}=1$: 
    (a) Dispersion of the polariton energy $\hbar\omega_\pm$ (solid blue) and broadening $\hbar\gamma_\pm$ (dashed blue) as a function of the  normalized cooperative coupling parameter $\lambda\sqrt{N}/\omega_o$. (b): Comparison of the entanglement entropy due to the symmetrized output JSA with the entanglement entropy of the symmetrized input JSA presented in Fig.~\ref{Fig:empty_cav}(e).
    }
    \label{Fig:OccupCav}
\end{figure}

Moving beyond the cavity vacuum case to $\bar n \neq 0$, one must now use the full expressions provided by Eqs.~\eqref{JSA-c} and \eqref{JSA-TCr} to calculate the output JSA and associated entanglement entropy.  The key difference from the cavity vacuum case is the increased cavity dephasing rate $\gamma_c =\kappa[1/2+\bar{n}(\bar{n}+1)]$ due to the {\color{black} polariton-polariton  scattering } induced by the cavity photon population [Eq.~\eqref{gammac-def}]. Figure~\ref{Fig:OccupCav}(a) presents the dispersion relation for the case of $\bar{n}=1$. This results in a substantial increase of the cavity dephasing rate to $\gamma_c = 2.5\kappa = 62.5$~meV. For such strong broadening, the polariton states cannot enter the strong-coupling regime within the coupling range $0\leq\lambda\sqrt{N}\leq 0.01\omega_o$; note that the solid line in Fig.~\ref{Fig:OccupCav}(a) shows no Rabi splitting. Furthermore, the broadening exceeds the counter-diagonal width of the input JSA at $\lambda\sqrt{N} \sim 0$. As the coupling strength increases, the broadening decreases. This reduction in linewidth leads to a stronger overlap between the input JSA and the resonances of the biphoton scattering amplitudes, thereby enhancing the features in the output JSA. This behavior explains the rise in the entanglement entropy with increasing coupling strength, as observed in Fig.~\ref{Fig:OccupCav}(b). Nevertheless, a small change in the spectral overlap results in a slight increase in the entropy.

\begin{figure}[t]
\includegraphics[width=0.45\textwidth]{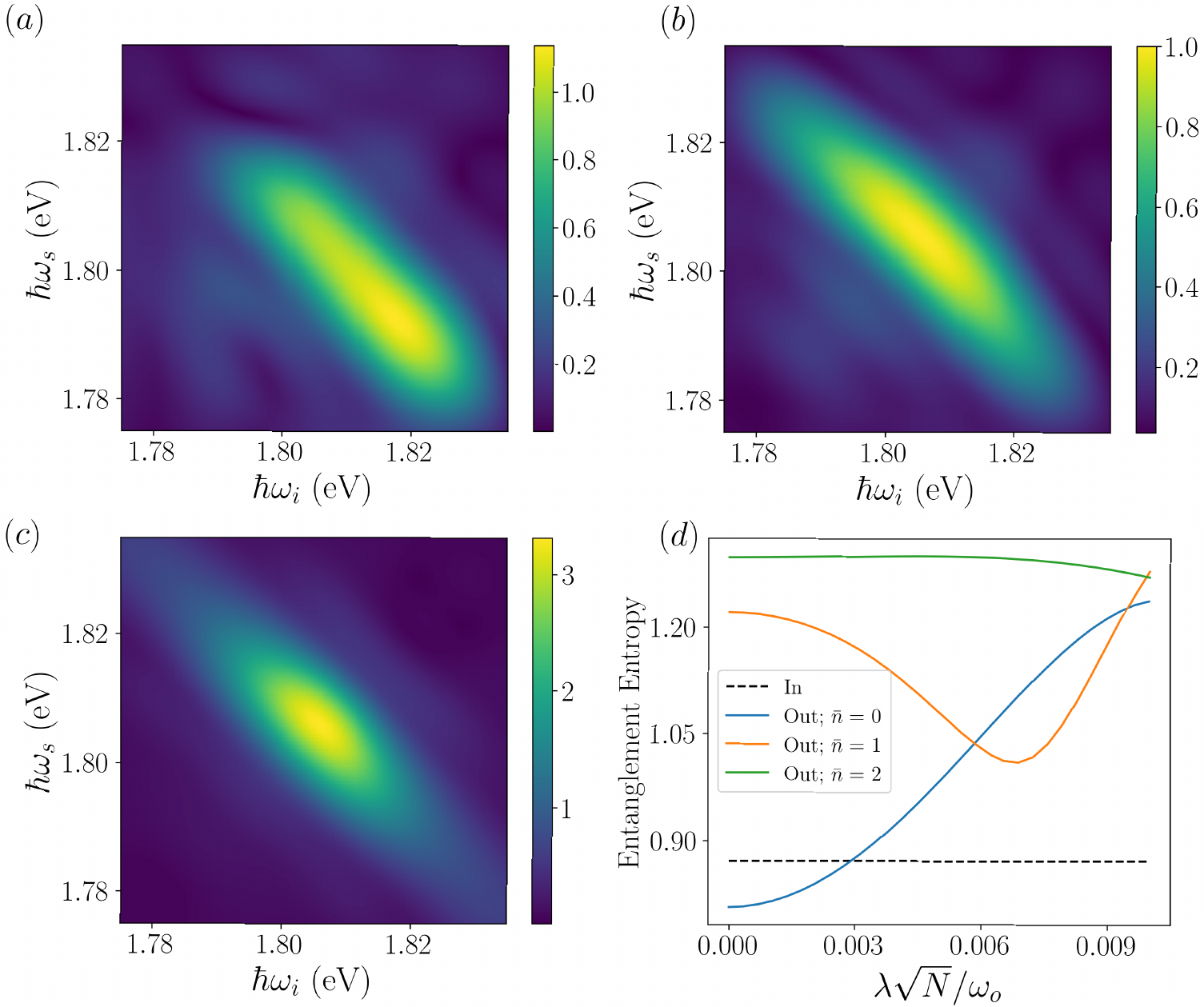}
    \caption{(a) Absolute value of input JSA obtained from experimental measurements of SPDC reported in Refs.~\cite{Moretti_JCP:2023, Malatesta_arXiv:2023}. (b) Absolute value of the symmetrized version for the input JSA from panel~(a). (c) Absolute value of symmetrized output JSA calculated using the input JSA for the empty cavity, $\lambda\sqrt{N}=\bar{n}=0$. (d) Comparison of entanglement entropy due to the input JSA presented in panel~(b) and symmetrized output JSA, for $\bar n = 0, 1,2$ vs. the normalized coupling strength. 
    }
    \label{Fig:ExpJSA}
\end{figure}

\subsection{Experimentally measured input JSA}

To complete our analysis, we consider biphoton scattering by the cavity using the experimentally measured input JSA shown in Fig.~\ref{Fig:ExpJSA}(a). In this case, the sum-over-polariton-pole expression [Eq.~\eqref{JSA-TCr}] cannot be applied to calculate the incoherent redistribution component, and we instead employ the general form for this contribution given in Eqs.~\eqref{JSA-r}. The coherent part is calculated using Eq.~\eqref{JSA-c}. To maintain consistency with the analysis above, we symmetrize the input JSA, as presented in Fig.~\ref{Fig:ExpJSA}(b). Compared to the model input JSA in Fig.~\ref{Fig:empty_cav}(e), the experimental JSA has shorter extent along the counter-diagonal. The calculated output JSA for the empty-cavity case is shown in Fig.~\ref{Fig:ExpJSA}(c) and reveals a narrowing along the counter-diagonal direction relative to the input JSA [Fig.~\ref{Fig:ExpJSA}(b)]. This behavior reflects the same spectral-filtering trend observed previously for the empty-cavity case [Figs.~\ref{Fig:empty_cav}(e) and \ref{Fig:empty_cav}(f)]. Accordingly, the increase in the associated entanglement entropy (blue curve in Fig.~\ref{Fig:ExpJSA}(d)) for the vacuum cavity can be attributed to the same spectral-filtering mechanism used to interpret Fig.~\ref{Fig:StongCoup}(c).

Considering the cavity contains $\bar{n}=1$ and $\bar{n}=2$ photons, the entanglement entropy presented in Fig.~\ref{Fig:ExpJSA}(d) does not exhibit significant variation compared to the empty cavity case. This behavior is similar to what we observed in the weak coupling regime shown in Fig.~\ref{Fig:OccupCav}{\color{black} (b)}, which can be interpreted as resulting from slow changes in the dephasing rate. Interestingly, we observe a non-monotonic behavior of the entanglement entropy for $\bar{n}=1$, with a minimum at $\lambda \sim 0.008\omega_o$, possibly arising from the interplay between entropy loss and production facilitated by the spectral properties. As the length of the input JSA along the counterdiagonal becomes larger (going from experimental to model input), this minimum shifts toward weaker couplings, leading to a small plateau observed in Fig.~\ref{Fig:OccupCav}(b) for the case of the model input JSA.  

\section{Discussion and Perspectives}
\label{Sec:Conc}

The variations in the biphoton entanglement entropy quantify the changes in biphoton quantum correlations. In the presence of uncorrelated polariton fluctuations, these correlations are diminished through dephasing processes, which consequently reduce the entanglement entropy~\cite{Li_JCP:2019}. Conversely, double excitations within the material system tend to generate entanglement entropy via cascaded biphoton processes~\cite{Bittner_JCP:2020}. In our case, the correlated bipolariton processes are incorporated into the scattering amplitude that governs the redistribution dynamics through the bipolariton Green functions. These contributions facilitate the production of entanglement entropy. Moreover, including polariton–polariton interactions through the vertex part of the biphoton Green function provides an additional mechanism for generating entanglement entropy~\cite{Bittner_JCP:2025}.

Our analysis in Sec.~\ref{Sec:Res} demonstrated that the entanglement entropy of the outgoing biphoton state arises not only from the interplay between decoherence processes and bipolariton correlations that lead to entropy production, but also depends on the input photon entanglement entropy encoded in the biphoton JSA. Notably, Figs.~\ref{Fig:OccupCav}{\color{black} (b)} and \ref{Fig:ExpJSA}(d) reveal distinct variation in magnitudes for biphoton entanglement entropy after scattering by the cavity in vacuum state when probed with biphoton pairs of high entanglement and lower entanglement. This distinction must be taken into account when employing entangled biphoton states to probe bipolariton correlations. Furthermore, isolating these correlations requires subtracting the contribution from coherent biphoton scattering. Since our analytical formulation explicitly separates these terms, the present theory provides a clear prescription for modeling and subsequently removing the coherent component while analyzing experimental data.

We showed that the outgoing biphoton entanglement entropy in the case of a cavity containing a steady-state photon population exhibits a more complex behavior with a relatively small variation range [Figs.~\ref{Fig:OccupCav}{\color{black} (b)} and \ref{Fig:ExpJSA}(d)]. Accordingly, we argue that biphoton scattering quantum light spectroscopy serves as a sensitive probe of polariton states in the photon-vacuum state. As the coupling strength increases, the cavity may transition into the ultrastrong coupling regime, where nontrivial correlations emerge between the photon vacuum and the QEs~\cite{fornRMP:2019,KockumNRP:2019,leBoiteAQT:2020}. In this regime, biphoton scattering can provide a powerful probe for these nontrivial correlations and their effect on the bipolariton interactions. However, such an application requires an input biphoton pair with a sufficiently broad JSA to span the spectral range encompassing the large Rabi splitting of the polariton states. Compared to conventional ultrafast techniques, the proposed biphoton scattering spectroscopy has advantage of using low photon excitation regime for probing the elementary excitations without affecting interactions between them and potentially the vacuum fluctuations. 

In conclusion, we have developed a tractable scattering theory framework for describing the scattering of a biphoton wavepacket through a photonic cavity supporting polariton and bipolariton quasiparticles. {\color{black} The theory advances the uncorrelated input biphoton scattering formalism, used in waveguide cavity QED~\cite{Sheremet_RMP:2023}, to the case of entangled photon pairs, opening new perspectives for its applications.} The adopted Green function representation provides a versatile approach for modeling quasiparticle dynamics in molecular and semiconductor materials coupled to a cavity mode. Our model calculations based on the TC model revealed that the output JSA is a result of an interplay between polariton and bipolariton Fano-type resonance and the input JSA. Moreover, we identified the limitations imposed by the structure of the input JSA and demonstrated the sensitivity of the technique for probing polaritons supported by a cavity in the vacuum state. 

{\color{black} The formalism indeed enables modeling across a broad range of light–matter coupling regimes, from weak to strong and ultrastrong coupling, thereby opening a wide range of opportunities for further studies. The minimal modification required to incorporate this regime into our framework is the extension of the Tavis–Cummings model to the Dicke model by including counter-rotating terms in the cavity mode–QE interaction. While the Dicke model can provide an adequate description of cold atoms, molecular systems, and semiconductor quantum dots in microcavities, ultrastrong light–matter coupling has also been experimentally reported in van der Waals materials such as CrSBr, which host unique self-hybridized exciton–polariton states~\cite{CanalesJCP:2021,RutaNatComm:2023,DirnbergerNature:2023,WangNatCom:2023,Li_AdvFunMat:2024,Adak_AdvMater:2025}. This class of quantum materials therefore constitutes a highly attractive solid-state platform in which the strongly correlated nature of excitonic states and their coupling to other degrees of freedom can be probed using quantum light. In this case, a microscopic description of polaritonic states must go beyond the Dicke model.}

\begin{acknowledgments}
A.P.'s research was supported by the Laboratory Directed Research and Development (LDRD) program of Los Alamos National Laboratory under project number 20230347ER. N.J.'s work at LANL was supported by the U.S. Department of Energy Office of Science as part of the CHIME, a Microelectronics Science Research Center (MSRC). The work at the University of Houston, E.R.B, was supported by the National Science Foundation under CHE-2404788 and S.D. by the Robert A. Welch Foundation (E-1337). Y.Z. acknowledges the support from the US DOE, Office of Science, Basic Energy Sciences, Chemical Sciences, Geosciences, and Biosciences Division under Triad National Security, LLC (``Triad") contract Grant 89233218CNA000001 (FWP: LANLECF7). ARSK acknowledges funding from the National Science Foundation CAREER grant (CHE-2338663), This work was performed, in part, at the Center for Integrated Nanotechnologies, an Office of Science User Facility operated for the U.S. Department of Energy (DOE) Office of Science. Los Alamos National Laboratory, an affirmative action equal opportunity employer, is managed by Triad National Security, LLC for the U.S. Department of Energy’s NNSA, under contract 89233218CNA000001.
\end{acknowledgments}

\section*{Author Contribution Statement}
A.P., E.R.B., and A.R.S.K. jointly conceived the ideas, developed the general concepts, and supervised the project. A.P.,  S.D., and E.R.B. developed the theoretical framework with help from N.J. and Y.Z.; N.J. performed the model analysis with assistance from S.D. and A.R.S.K.; A.P. and N.J. prepared the initial draft of the manuscript, and all authors contributed to its final version. All authors discussed the results and approved the manuscript.

\section*{Conflicts of Interest}
The authors have no conflicts of interest to declare.


\appendix

{\color{black}
\section{Truncated power-series representation of the scattering matrix in Eq.~\eqref{scat-sef}}
\label{App:S-mtx}

In this Appendix, we provide details of the derivation of the scattering matrix components that lead to Eqs.~\eqref{scat-1} and \eqref{scat-2}.

We begin by expanding the time-ordered exponential in Eq.~\eqref{scat-sef} into a power series, truncated at the $(-i/\hbar)^4$ term.
\begin{widetext}
\begin{eqnarray}
\label{scat-power-series}
\hat S &\approx& 1 -\frac{i}{\hbar}\int\limits_{-\infty}^\infty dt_1 ~\left\langle\hat H_\texttt{I} (t_1)  \right\rangle
    +\left(\frac{-i}{\hbar}\right)^2\int\limits_{-\infty}^\infty dt_1\int\limits_{-\infty}^\infty dt_2 \theta(t_1-t_2) ~\left\langle\hat H_\texttt{I} (t_1) \hat H_\texttt{I} (t_2)  \right\rangle
\\\nonumber &+& 
\left(\frac{-i}{\hbar}\right)^3\int\limits_{-\infty}^\infty dt_1\int\limits_{-\infty}^\infty dt_2 
\int\limits_{-\infty}^\infty dt_3\theta(t_1-t_2)\theta(t_2-t_3) 
~\left\langle\hat H_\texttt{I} (t_1) \hat H_\texttt{I} (t_2) \hat H_\texttt{I} (t_3)  \right\rangle
\\\nonumber &+& 
\left(\frac{-i}{\hbar}\right)^4\int\limits_{-\infty}^\infty dt_1\int\limits_{-\infty}^\infty dt_2 
\int\limits_{-\infty}^\infty dt_3\int\limits_{-\infty}^\infty dt_4\theta(t_1-t_2)\theta(t_2-t_3)\theta(t_3-t_4) 
~\left\langle\hat H_\texttt{I} (t_1) \hat H_\texttt{I} (t_2) \hat H_\texttt{I} (t_3)\hat H_\texttt{I} (t_4)  \right\rangle.
\end{eqnarray}

By substituting the cavity photon–photon continuum interaction Hamiltonian~\eqref{H-ph-int} into Eq.~\eqref{scat-power-series}, we obtain
\begin{eqnarray}
\label{scat-power-series-1}
\hat S &\approx& 1 -\kappa^{1/2}\int d\omega_1\int\limits dt_1 
    ~\left\langle\left[\ha^\dag (t_1)\hb(\omega_1)e^{-i\omega_1 t_1}
        -\hb^\dag(\omega_1)\ha (t_1)e^{i\omega_1 t_1}\right]  \right\rangle
\\\nonumber&+&
    \kappa \iint d\omega_1 d\omega_2\iint dt_1 dt_2 
    \theta(t_1-t_2) 
    ~\left\langle\left[\ha^\dag (t_1)\hb(\omega_1)e^{-i\omega_1 t_1}
        -\hb^\dag(\omega_1)\ha (t_1)e^{i\omega_1 t_1}\right] 
        \left[\ha^\dag (t_2)\hb(\omega_2)e^{-i\omega_2 t_2}
        -\hb^\dag(\omega_2)\ha (t_2)e^{i\omega_2 t_2}\right] \right\rangle
\\\nonumber &-& 
\kappa^{3/2} \iiint d\omega_1 d\omega_2 d\omega_3\iiint dt_1 dt_2 dt_3 
    \theta(t_1-t_2) \theta(t_2-t_3) 
    ~\left\langle\left[\ha^\dag (t_1)\hb(\omega_1)e^{-i\omega_1 t_1}
        -\hb^\dag(\omega_1)\ha (t_1)e^{i\omega_1 t_1}\right]
\right.\\\nonumber &\times&\left.        
        \left[\ha^\dag (t_2)\hb(\omega_2)e^{-i\omega_2 t_2}
        -\hb^\dag(\omega_2)\ha (t_2)e^{i\omega_2 t_2}\right]
        \left[\ha^\dag (t_3)\hb(\omega_3)e^{-i\omega_3 t_3}
        -\hb^\dag(\omega_3)\ha (t_3)e^{i\omega_3 t_3}\right]\right\rangle
\\\nonumber &+& 
\kappa^{2} \iiiint d\omega_1 d\omega_2 d\omega_3 d\omega_4\iiiint dt_1 dt_2 dt_3 dt_4 
    \theta(t_1-t_2) \theta(t_2-t_3) \theta(t_3-t_4) 
\\\nonumber&\times&    
    ~\left\langle\left[\ha^\dag (t_1)\hb(\omega_1)e^{-i\omega_1 t_1}
        -\hb^\dag(\omega_1)\ha (t_1)e^{i\omega_1 t_1}\right]
        \left[\ha^\dag (t_2)\hb(\omega_2)e^{-i\omega_2 t_2}
        -\hb^\dag(\omega_2)\ha (t_2)e^{i\omega_2 t_2}\right]
\right.\\\nonumber &\times&\left.        
        ~~\left[\ha^\dag (t_3)\hb(\omega_3)e^{-i\omega_3 t_3}
        -\hb^\dag(\omega_3)\ha (t_3)e^{i\omega_3 t_3}\right]
        \left[\ha^\dag (t_4)\hb(\omega_4)e^{-i\omega_4 t_4}
        -\hb^\dag(\omega_4)\ha (t_4)e^{i\omega_4 t_4}\right]\right\rangle,
\end{eqnarray}
\end{widetext}
where, for the sake of brevity, we have omitted the $\pm\infty$ integration limits.

Given that we consider photon scattering processes that conserve the number of photons entering and exiting the cavity, in Eq.~\eqref{scat-power-series-1} we retain only the terms containing equal numbers of photon continuum creation operators, $\hat b^\dag(\omega)$, and annihilation operators, $\hat b(\omega)$. The terms of order $\kappa^{1/2}$ and $\kappa^{3/2}$, by construction, contain an odd number of photon continuum operators and must therefore be excluded. Further expanding the operator products in the terms prefactored by $\kappa$ and $\kappa^2$, and bringing the photon continuum operators into normal order, we retain the terms containing the operator products $\hat b^\dag(\omega)\hat b(\omega')$ and $\hat b^\dag(\omega)\hat b^\dag(\omega')\hat b(\omega'')\hat b(\omega''')$, respectively. This procedure leads directly to Eqs.~\eqref{scat-1} and \eqref{scat-2}. The terms arising from contractions of the photon continuum operators during normal ordering correspond to the single-polariton self-energy, which is derived in the next Appendix~\ref{App:corr} and discussed in Sec.~\ref{Sec:S-matrix}.

}

\section{{\color{black} Evaluation  of self-energy term in polariton Green function and associated dephasing rate}}
\label{App:corr}

The non-zero contractions, which represent the difference between actual and normal ordering, of the photon continuum operator pairs in the second and third terms of Eq.~\eqref{scat-2} lead to additional contributions to the scattering operator. In this appendix, we evaluate these contributions and demonstrate that they provide the leading corrections to the {\color{black} self-energy term} in polariton Green function. 

Specifically, we aim to isolate the terms involving normally ordered pairs of the photon operators. These terms can be derived from Feynman diagrams (b, c) and (e, f) in Fig.~\ref{Fig:FD1} by connecting the incident and scattered photon lines representing the anti-ordered operators. We examine four contractions as shown in Fig.~\ref{Fig:FD1-corr}. The remaining two contractions of the $\omega_1$ and $\omega_2$ photon lines in diagrams (c) and (f) of Fig.~\ref{Fig:FD1} contribute to the loops attached to photon emission vertices, leading to vertex renormalization. The first loop, indeed, produces divergent terms. {\color{black} We exclude the vertex corrections from our analysis for the reason they do not contribute to the dephasing rate.}

The expressions for the four terms forming the scattering operator obtained from diagrams (a)-(d) of Fig.~\ref{Fig:FD1-corr} are given by
\begin{eqnarray}
\label{S-corr}
\hspace{-0.5cm}\hat S^{(2)}_\text{corr}  
&=& -2\pi\kappa^2\int d\omega\int d\omega'
\left\{
\frac{i\left[G_{}(\omega)-G^*_{}(\omega')\right]G_{}(\omega)}{\omega-\omega'+i0}
\right.\\\nonumber&+&\left.
\left[G_{}(\omega)-G^*_{}(\omega')\right]\bar{\cal G}(\omega-\omega')G_{}(\omega)
\right.\\\nonumber &+&\left. 
c.c.\right\}\hb^\dag(\omega)\hb(\omega),
\end{eqnarray}
respectively. Here c.c. represents the complex conjugate terms.

By combining Eq.~\eqref{S-corr} with Eq.~\eqref{scatGF-1}, which describes the single-photon scattering operator, we can derive the following representation
\begin{eqnarray}
\nonumber
\hat{S}^{(1)}+\hat S^{(2)}_\text{corr}  &=& - 2\pi i \kappa \int d\omega G_{}(\omega)\left[1+\Sigma(\omega)G_{}(\omega)\right] 
\\\label{scatGF-dressed} 
&~& \times
:\hb^\dag(\omega) \hb(\omega):
+h.c., 
\end{eqnarray}
where we identify the self-energy term
\begin{eqnarray}
\label{SelfE-def}
\Sigma(\omega) &=& i\kappa\int d\omega' \left[1-\frac{G^*_{}(\omega')}{G_{}(\omega)}\right]
\\\nonumber&~&\times
\left[\frac{i}{\omega-\omega'+i0}+\bar{\cal G}(\omega-\omega')
\right].
\end{eqnarray}

To calculate the integral over $d\omega'$, we use the Green functions represented by Eqs.~\eqref{Gapls} -- \eqref{upm-cc} and \eqref{Gdc-12}. The result of the calculation is
\begin{eqnarray}
\label{SelfE-part}
\Sigma(\omega) &=& 2\pi i \gamma_c+\Delta\Sigma(\omega),
\end{eqnarray}
where we isolated the frequency independent {\color{black} cavity dephasing rate} 
\begin{eqnarray}
\label{gammac-def}
\gamma_c = \kappa\left[\frac{1}{2}+\bar{n}\left(\bar n +1\right)\right] ,
\end{eqnarray}
and frequency dependent self-energy correction
\begin{eqnarray}
\label{DSF-omg}
\Delta\Sigma(\omega)&=& 4\pi i \kappa\frac{G^*_{}(\omega)}{G_{}(\omega)}.
\end{eqnarray}

We accounted for the frequency independent component given by Eq.~\eqref{gammac-def} by including it directly into the cavity dephasing rate in Eqs.~\eqref{Ga-EOM} --\eqref{Gs-EOM-FD} for the Green function. According to Eq.~\eqref{scatGF-dressed}, the other correction provided by Eq.~\eqref{DSF-omg}) can be directly accounted for in the coherent scattering amplitude in Eq.~\eqref{S2ph-c} by multiplying the Green functions there with the $\left[1+4\pi i\kappa G^*_{}(\omega)\right]$ factor and its complex conjugate.  

\section{Polariton Green function in Tavis-Cummings model}
\label{Appx:TCModel}

Starting with the TC Hamiltonian provided by Eq.~~\eqref{H-TC-def}, we derive the equations of motion for the polariton Green function introduced in Eq.~\eqref{G2t-1}. By differentiating this Green function with respect to the first time argument and applying the Heisenberg equation of motion to the time derivative of the cavity mode operator, we arrive at the following equations of motion for the Green function
\begin{eqnarray}
\label{Gt-EOM}
    i\partial_t G(t_1-t_2)&=&\delta(t_1-t_2)\left\langle \ha(t_2)\ha^\dag(t_2)\right\rangle
\\\nonumber&-&
    i\theta(t_1-t_2)\left\langle \left[\ha,\hat H_\text{TC}\right](t_1)\ha^\dag(t_2)\right\rangle.
\end{eqnarray}
This Green function satisfies the initial condition at $t_1=t_2$ defined by the time correlation function $\left\langle \ha(t_2)\ha^\dag(t_2)\right\rangle$, which describes the time evolution of the cavity mode population at $t_2$. In Sec.~\ref{Sec:S-matrix}, we restricted our analysis to the scenario in which the cavity mode and the QEs reach a steady-state regime characterized by time-independent populations $\bar n+1=\left\langle \ha\ha^\dag\right\rangle$ for the cavity mode and  $\bar\sigma_z=2\sum_{n=1}^N\left\langle \hsp_n\hsm_n\right\rangle/(N-1)$ for QEs. 

To obtain a closed set of equations for the polariton Green function, we complement $G_{}(\omega)$ with another polariton Green function, $\bar{G}(t)=-i\theta(t)\sum_{n=1}^N\left\langle \hsm_n(t)\ha^\dag(0)\right\rangle/N$, which describes the coherence between the QEs and the cavity mode. The initial condition for this Green function is given by the simultaneous coherence between the QE-cavity mode, denoted as $ \overline{\sigma a}\equiv\sum_{n=1}^N\left\langle \hsm_n\ha^\dag\right\rangle/N$, which is also present in the steady state. 

Utilizing Eq.~\eqref{Gt-EOM} with the TC Hamiltonian~\eqref{H-TC-def} for the Green functions $G(\omega)$ and $\bar G(\omega)$, we obtain the following closed set of equations of motion
\begin{eqnarray}
\label{Ga-EOM}
  &~& \left( i\partial_t-\tilde{\omega}_c \right) G(t)-\lambda N \bar G(t)=\delta(t)\left(\bar{n}+1\right)
\\\label{Gs-EOM}
   &~& \left( i\partial_t-\tilde{\omega}_o \right) \bar G(t)+\lambda \bar \sigma_z N  G(t)=\delta(t)\overline{\sigma a}.
\end{eqnarray}
Here, we introduce the shorthand notations for the complex cavity, $\tilde\omega_c = \omega_c - i\gamma_c$, and QE, $\tilde\omega_o = \omega_o - i\gamma_o$, frequencies. The cavity dephasing rate, $\gamma_c$, includes a contribution of $\kappa/2$ and additional radiative corrections, as derived in Appendix~\ref{App:corr}. 

After transforming Eqs.~\eqref{Ga-EOM} and \eqref{Gs-EOM} into the Fourier domain 
\begin{eqnarray}
\label{Ga-EOM-FD}
  &~& \left( \omega-\tilde{\omega}_c \right) G(\omega)-\lambda N \bar G(\omega)=\left(\bar{n}+1\right)
\\\label{Gs-EOM-FD}
   &~& \left( \omega-\tilde{\omega}_o \right) \bar G(\omega)+\lambda \bar \sigma_z N  G(\omega)=\overline{\sigma a},
\end{eqnarray}
one can easily solve them. The solutions for the Green function and its complex conjugate read
{\color{black}
\begin{eqnarray}
\label{Gsc-def}
G_{}(\omega)&=&\frac{\left(\omega-\tilde\omega_o\right)\left(\bar{n}+1\right)+\lambda N \overline{\sigma a}}
    {2\pi\left[(\omega -\tilde\omega_c)(\omega-\tilde\omega_o)
       +\lambda^2 N\bsz\right]}.
\\\label{Gsc-cc}
G^*_{}(\omega)&=&\frac{\left(\omega-\tilde\omega^*_o\right)\bar{n}+\lambda N \overline{\sigma a}^*}
    {2\pi\left[(\omega -\tilde\omega^*_c)(\omega-\tilde\omega^*_o)
       +\lambda^2 N\bsz\right]},
\end{eqnarray}
}
where the $2\pi$ factor provides the Fourier integral normalization. These expressions correspond to the representations provided in Eqs.~\eqref{Gapls}--\eqref{upm-cc}. To obtain the pole expansion, we evaluate the residues of the Green functions in Eqs.~\eqref{Gsc-def} and \eqref{Gsc-cc} at each pole [Eq.~\eqref{wpm-def}], giving rise to the expansion coefficients specified in Eqs.~\eqref{upm-def}--\eqref{upm-cc}, respectively.

\section{Analysis of Schmidt coefficient distribution}
\label{App:Schidt}

Consider a JSA that only has non-zero amplitude close to the counter-diagonal $\omega_s + \omega_i = \omega_p$ (such as the input and output JSAs in Figs. \ref{Fig:empty_cav} and \ref{Fig:StongCoup}). As the width of the JSA perpendicular to the counter-diagonal tends to 0, the JSA approaches the Schmidt decomposed form $\mathcal{F}(\omega_s, \omega_i) \propto \int d\omega \; r(\omega) \delta(\omega_s - \omega)\delta(\omega_i +\omega - \omega_p)$. Here, the Schmidt coefficient $r(\omega)$ governs the broadness of the JSA along the counter-diagonal. A narrow JSA corresponds to a sharply peaked $r(\omega)$, with the limit $r(\omega) \to \delta(\omega)$ analogous to an unentangled state with only a single non-zero Schmidt coefficient. On the other hand, a broader JSA corresponds to a widened profile of $r(\omega)$ with a weaker dependence on $\omega$; the limit of $r(\omega)$ approaching a constant independent of $\omega$ corresponds to a maximally entangled state (i.e. partial tracing one of the photon modes would yield a density operator proportional to the identity). Therefore, the output JSAs in the main text become more entangled as they widen and approach a maximally entangled state.


\bibliography{bib-local}
\end{document}